\documentclass[aps,prd,reprint,showpacs,preprintnumbers,superscriptaddress,nofootinbib]{revtex4-1}
\usepackage{slashed}
\usepackage{bm} 
\usepackage{epsf}
\usepackage{graphicx,epsfig}
\usepackage{latexsym,amssymb,amsmath,float,url}
\usepackage{bbold}
\usepackage{latexsym}

\input{epsf}

\newcommand\nubar{{\bar\nu}}
\newcommand{\Psibar}{{\bar \Psi}}
\newcommand{\photino}{{\tilde \gamma}}

\def\bs{\begin{split}}
\def\es{\end{split}}

\def\e{{\rm e}}

\def\5{\overline 5}

\renewcommand{\)}{\right)}

\newcommand{\Order}{\mathcal{O}} 

\newcommand{\half}{\frac{1}{2}}
\newcommand{\quarter}{\frac{1}{4}}

\newcommand{\beq}[1]{\begin{equation}\label{#1}}
\newcommand{\eeq}{\end{equation}}
\newcommand{\bea}[1]{\begin{eqnarray}\label{#1}}
\newcommand{\eea}{\end{eqnarray}}
\newcommand{\unitmatrix}{\openone}
\newcommand{\rf}[1]{(\ref{#1})}
\newcommand{\barr}{\begin{array}}
\newcommand{\earr}{\end{array}}
\newcommand{\rarr}{\rightarrow}
\newcommand{\longrarr}{\longrightarrow}

\newcommand{\bsigma}{\overline\sigma}

\def\be{\begin{equation}}
\def\ee{\end{equation}}
\def\ba{\begin{eqnarray}}
\def\ea{\end{eqnarray}}

\newcommand{\X}{{\rm X}}
\newcommand{\Y}{{\rm Y}}
\newcommand{\Ghat}{{\hat\Gamma}}
\newcommand{\tS}{{\tilde S}}
\newcommand{\tP}{{\tilde P}}
\newcommand{\tV}{{\tilde V}}
\newcommand{\tA}{{\tilde A}}
\newcommand{\tT}{{\tilde T}}
\newcommand{\hT}{{\hat T}}

\newcommand{\ubar}{{\overline u}}
\newcommand{\vbar}{{\overline v}}
\newcommand{\chibar}{{\overline\chi}}
\newcommand{\ellbar}{{\overline\ell}}

\begin{document}



\title{W/Z Bremsstrahlung as the Dominant Annihilation Channel for Dark Matter}

\author{Nicole F.\ Bell} 
\affiliation{School of Physics, The University of Melbourne, 
Victoria 3010, Australia}

\author{James B.\ Dent}
\affiliation{Department of Physics and School of Earth and Space Exploration,
Arizona State University, Tempe, AZ 85287-1404, USA}

\author{Thomas D.\ Jacques}
\affiliation{School of Physics, The University of Melbourne, 
Victoria 3010, Australia}

\author{Thomas J.\ Weiler}
\affiliation{Department of Physics and Astronomy,
Vanderbilt University, Nashville, TN 37235, USA}

\date{September 23, 2011}

\begin{abstract}

Dark matter annihilation to leptons, $\chi\chi \rightarrow \ell\ellbar$, is
necessarily accompanied by electroweak radiative corrections, in which
a $W$ or $Z$ boson is radiated from a final state particle.
Given that the $W$ and $Z$ gauge bosons decay dominantly via
hadronic channels, it is thus impossible to produce final state
leptons without accompanying protons, antiprotons, and gamma rays.   
Significantly, while many dark matter models feature a helicity suppressed
annihilation rate to fermions, radiating a massive gauge boson from a
final state fermion removes this helicity suppression, such that the
branching ratios Br($\ell \nu W $), Br($\ell^+\ell^-Z$), and Br($\nu\nubar Z$)
dominate over Br($\ell\ellbar$).  
$W/Z$-bremsstrahlung thus allows indirect detection of many WIMP models
that would otherwise be helicity-suppressed, or $v^2$ suppressed.
Antiprotons and even anti-deuterons become consequential final state particles.
This is an important result for future DM searches.
We discuss the implications of $W/Z$-bremsstrahlung for ``leptonic'' DM
models which aim to fit recent cosmic ray positron and antiproton data.

\end{abstract}

\pacs{95.35.+d, 12.15.Lk, 95.85.Ry}


\onecolumngrid
\section*{Erratum}
\label{sec:erratum}

Due to an error in a Fierz identity published elsewhere~\cite{Okun:1982ap}, some
of the results presented in this paper are also in error.  
Here we outline which results of this paper are correct and which are incorrect, 
and then briefly discuss the consequences for the latter.  
Further details can be found in~\cite{revisited}.

\begin{figure}[t]
\includegraphics[width=1.0\columnwidth]{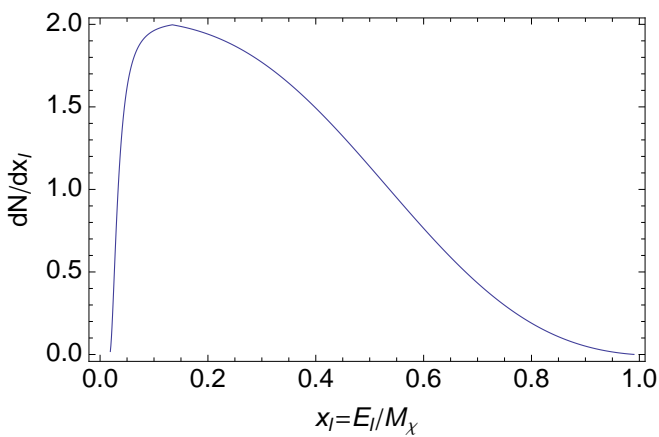}
\caption{
\label{fig:SecLepSpectrum}
The secondary lepton spectrum (i.e., from $W\rarr \nu_\ell \ell$)
per $W$ for the example model, 
with $M_\chi=300$ GeV and $M_\eta^2\gg M_\chi^2$. (The branching ratio
for $W\rarr \nu l$, 11\% per flavor, is not included here.)
}
\end{figure}

The various Fierz identities we presented in this paper (Eqns.~(\ref{chiralFierz1},\ref{chiralFierz},\ref{genFierz})
and all the equations of Appendix \ref{app:Fierz}) are all correct.  	
However, due to parallel-processing of our efforts, our 
explicit cross section calculation was performed using the Fierz
identity given in Okun's textbook.  
This identity should correctly read

\begin{equation}
\label{FierzUpdated}
F_l^i G_k^m = \frac{1}{4}\sum_A \Delta_A (F \gamma_A G)_l^m (\gamma_A)_k^i,
\end{equation}
in correspondence with our Eq.~(\ref{XandY1}), but does not.  
In the published expression of Ref~\cite{Okun:1982ap}, 
the indices $\{k,i\}$ are incorrectly interchanged with $\{l,m\}$ on one side of the equation, 
which is equivalent to exchanging $F$ and $G$ on one side of Eq.~(\ref{FierzUpdated}).
This transposition is thus not an issue for the usual application of the Fierz
identity to $2\rightarrow 2$ processes, where $F = G$.
However, it becomes an issue for $2\rightarrow 3$ processes. 
As a consequence, the part of our paper that made use of Okun's textbook formula contains errors.
The Fierz'ed matrix elements of Eqns.~(\ref{MA}--\ref{MD}) are incorrect, as are
the results that follow from them, namely, Eqns.~(\ref{MatElement}, \ref{vsigma}, \ref{dsigdEW}), 
Eqns.~(\ref{vsigmaFull}--\ref{dsigmadEl}), Eqns.~(\ref{ellipse1} -- \ref{ellipse3}) and Figures \ref{ratio1},\ref{dNdEW},\ref{dNdEe},\ref{fig:SecLepSpectrum}.

The correct results are that 
the s-wave contributions from the four matrix elements of Eqns.~(\ref{MA}--\ref{MD}) cancel exactly in the four
Fermi limit where $M_\eta \gg M_\chi$.  
Since the four Fermi limit gives a zero s-wave result, there are no interesting new figures to be 
presented in this erratum.
However, if the four Fermi limit is not
adopted, then the cancellation is not complete and an unsuppressed s-wave amplitude results.
This means that the interesting parameter space is that where 
$M_\eta$ and $M_\chi$ are comparable.  
In this regime, one should consider not only the 4
diagrams of Fig.~\ref{feyngraph1}, but also the two additional diagrams in which the
$W/Z$ boson is radiated from the $\eta$ propagator.  
This in turn means that the leading s-wave amplitude is one power of $(M_\chi/M_\eta)^2$ higher 
than implied by our errant calculation.
In Ref.~\cite{revisited}, with two additional authors, we present the higher order calculation
of the unsuppressed s-wave cross section.

Many of the important qualitative conclusions of this paper still hold:

(i) $W$/$Z$ bremsstrahlung can lift helicity suppressions and thus be the
dominant annihilation model, albeit for a region of parameter space smaller than 
we originally proposed.

(ii) antiprotons produced by the decays of the $W$ and $Z$ gauge bosons
can prove lethal for models attempting to produce positrons without
overproducing antiprotons.

We thank Ahmad Galea for his signifiant help in identifying the errant formulas.

\newpage
\twocolumngrid
\maketitle

\section{Introduction}
\label{sec:intro}

\noindent
An abundance of cosmological and astrophysical evidence attests to the
existence of dark matter (DM), whose presence is inferred via its
gravitational
influence~\cite{Kamionkowski_review,Bertone_review,Bergstrom_review}.
However, the fundamental particle properties of DM remain essentially
unknown.  One important means of probing DM's particle nature is via
indirect detection, whereby we search for products of DM annihilation
(or decay) emanating from regions of DM concentration in the Universe
today.

The dark matter annihilation cross section is often parametrized as
$\langle v\sigma_A \rangle = a + bv^2 +\cdots$, where $\langle
v\sigma_A \rangle$ is the thermally-averaged annihilation cross
section.  The constant $a$ comes from s-wave annihilation, while the
velocity suppressed $bv^2$ term receives both s-wave and p-wave
contributions; the $L^{th}$ partial wave contribution to the
annihilation rate is suppressed as $v^{2L}$.  Given that $v \sim
10^{-3}c$ in galactic halos, even the p-wave contribution is highly
suppressed and thus only the s-wave contribution is expected to be
significant in the Universe today.  However, in many DM models the
s-wave annihilation into a fermion pair $\chi\chi \rightarrow
\bar{f}f$ is helicity suppressed by a factor $(m_f/M_\chi)^2$ (only
$\rightarrow \bar{t}t$ modes remain of interest, and then only for a
certain range of $\chi$~mass).

When computing DM annihilation signals, it is normally assumed that
only the lowest order tree-level processes make a significant
contribution.  However, there are important exceptions to this
statement.  Dark matter annihilation into charged particles, $\chi\chi
\rightarrow \bar{f}f$, is necessarily accompanied by the internal
bremsstrahlung process $\chi\chi \rightarrow \bar{f}f\gamma$, where
the photon may be radiated from one of the external particle legs
(final state radiation, FSR) or, possibly, from a virtual propagator
(virtual internal bremsstrahlung, VIB).  On the face of it, the
radiative rate is down by the usual QED coupling factor of
$\alpha/\pi\sim 500$.  However, and significantly, photon
bremsstrahlung can lift the helicity suppression of the $s$-wave
process~\cite{hardgamma}, which more than compensates for the extra
coupling factor.  Such a striking enhancement can arise when a
symmetry of the initial state $\chi\chi$ is satisfied by the three
body final state $\bar{f}f\gamma$, but not by the two body final state
$\bar{f}f$.  For bremsstrahlung of photons, only VIB is effective in
lifting the helicity suppression, as FSR is dominated by soft or
collinear photons (such that the two and three body final states have
the same symmetry properties) as discussed in
Ref. \cite{Bringmann:2007nk}.

In this paper we examine electroweak
bremsstrahlung~\cite{Berezinsky:2002hq,Kachelriess,BDJW,Dent:2008qy,Ciafaloni:2010qr,KSS09,Chen:1998},
i.e., bremsstrahlung of $Z$ or $W^\pm$ electroweak gauge bosons to
produce $\bar{f}f Z$ and $\bar{\ell}\nu W$ final states.  
The virtue for $W/Z$~bremsstrahlung to lift initial-state velocity and 
final-state helicity suppressions,
alluded to in~\cite{BDJW,KSS09}, has not been previously explored.
We show that
$W/Z$-bremsstrahlung can also lift suppression and become the dominant
annihilation channel.  Thus, $W/Z$~bremsstrahlung allows indirect
detection of many WIMP models that would otherwise be
helicity-suppressed, or $v^2$ suppressed.  This is an important result
for future DM searches.

There are a number of important distinctions between electromagnetic (EM)
and electroweak (EW) bremsstrahlung. An obvious one is that EM bremsstrahlung
produces just photons, whereas EW bremsstrahlung and subsequent decay
of the gauge bosons leads to leptons, hadrons and gamma rays, offering
correlated ``multi-messenger'' signals for indirect dark matter searches.
Another distinction is that $W/Z$-bremsstrahlung from final state particles (FSR) 
is sufficient to lift a suppression.  This is due to the nonzero gauge boson masses, and the
coupling of the gauge bosons to the non-conserved axial current which
leads to a different form for the polarization sum than in the case of
the photon (or a gluon in the similar QCD process). In contrast, for the EM process, 
VIB is required for the photon to lift a suppression.  Because an additional propagator 
appears for VIB, suppression-lifting EM bremsstrahlung is itself suppressed 
by an additional factor of $M^2_\chi/M^2_\eta$ relative to 
electroweak's FSR, where $M_\eta$ is the mass of the internal exchange-particle.
Only in the event of a near-degeneracy $M_\chi\sim M_\eta$ is this relative suppression of 
EM bremsstrahlung negligible.

DM annihilation to charged leptons has been the subject of much recent
attention, due to recently measured cosmic ray anomalies which point
to an excess of cosmic ray positrons above those that may be
attributed to conventional astrophysical processes.  PAMELA has
observed a sharp excess in the $e^+/(e^- + e^+)$ fraction at energies
beyond approximately 10 GeV~\cite{Pamela_positrons}, without a
corresponding excess in the antiproton/proton
data~\cite{Pamela_antiprotons,:2010rc}, while Fermi and HESS have
reported more modest excesses in the $(e^- + e^+)$ flux at energies of
order 1 TeV~\cite{Fermi1}.  These signals have led to a re-examination
of positron production in nearby pulsars~\cite{pulsars}, emission from supernova
remnants~\cite{supernova}, acceleration of $e^+e^-$ in cosmic ray
sources~\cite{accel}, and propagation in
conventional cosmic ray models~\cite{prop}.  As an
alternative to these astrophysical mechanisms, it has also been
proposed that the excess $e^+$ and $e^-$ are produced via dark matter
annihilation in the Galactic halo, with an abundance of DM models
proposed to accomplish this end.
A recent overview of $e^\pm$-excess data and possible interpretations is 
available in~\cite{Fan:2010yq}.

However, some of the most popular models suffer from helicity or
$v^2$-suppression.  A prototypical example of suppressed production of 
Standard Model (SM) fermion pairs is provided by
supersymmetry: Majorana neutralinos annihilate into a
pair of SM fermions via $t$- and $u$-channel exchange
of $SU(2)$-doublet sfermions.  To overcome the suppression, proponents
of these models have invoked large ``boost'' factors.  These boost
factors may be astrophysical in origin, as with postulated local
over-densities of dark matter, or they may arise from particle
physics, as with the Sommerfeld enhancement that arises from light
scalar exchange between dark matter particles.
Although not ruled out, these factors do seem to be a contrivance
designed to overcome the innate suppression.

A further problem with suppressed models is the
overproduction of antiprotons from unsuppressed $W/Z$ bremsstrahlung.
Given that hadronic decay modes of the $W$ and $Z$ bosons will lead to
significant numbers of both antiprotons and gamma rays, this will
impact the viability of models that might otherwise have explained the
observed positron excess.  Even in models which do not feature a
suppression, the W/Z-bremsstrahlung has important phenomenological
consequences, as the decay products of the gauge bosons make a pure
leptonic $e^+e^-$ signal impossible~\cite{KSS09}.

In Section~\ref{sec:Suppression} we discuss the circumstances under
which dark matter annihilation may be suppressed, and in
Section~\ref{sec:unsuppression} explain how $W/Z$ bremsstrahlung is
able to circumvent such a suppression.  In Section~\ref{sec:example}
we consider a representative model, and explicitly calculate the cross
sections for both the lowest order annihilation process, and for the
$W/Z$ bremsstrahlung process.  We discuss implications of these results
in Section~\ref{sec:conclusion}.  Calculational details are collected in 
five Appendices.  

\section{Understanding Suppression using Fierz Transformations}
\label{sec:Suppression}

In this section we describe the origin of $v^2$ and helicity
suppressions.  We shall make use of Fierz transformation and partial
wave decomposition to determine under what circumstances these
suppressions will or will not arise.

Dark matter candidates may be scalar, fermionic, or vector in nature;
if fermionic, they may be either Dirac or Majorana.  Permissible
annihilation models include $s$-, $t$-, and $u$-channel exchanges of a
new particle, and the various possibilities are listed in
Refs.~\cite{Lindner:2010rr,Randall,Kolb}.  In every case, it is useful
to classify the partial waves available to the decay process, and to
analyze the dependence on the mass of the SM particle-pair in the
final state.  In this article, we focus on fermionic Majorana dark
matter.

For fermionic dark matter, the natural projection of $2\rightarrow 2$
processes into partial waves makes use of the Fierz transformation.
In the next subsection we consider DM annihilation via the process
$\chi\chi \rightarrow \bar{f}f$, and explain the use of Fierz
transforms to convert the matrix elements for $t$/$u$-channel
annihilation, which are of the form $(\chibar \; \Gamma_A l)(\bar l \;
\Gamma_B \chi)$, to a sum of $s$-channel amplitudes of the form
$(\chibar \; \Gamma_1 \chi)(\bar l \; \Gamma_2 l)$.
In the following subsection we then categorize the Fierzed $s$-channel
amplitudes into partial waves and fermion-pair spin states, which
determines whether the amplitudes are velocity suppressed,
mass-suppressed, or unsuppressed.  In the third and final subsection,
we put our findings together to determine which class of models will
have a suppressed $2\rarr 2$ annihilation.  We show that in a certain
popular class of suppressed models, the $2\rarr 3$
$W/Z$-bremsstrahlung process is unsuppressed, and in fact dominant
for $2\,M_\chi > M_W$.  We will find in Section~\ref{sec:unsuppression} 
that a generalization of the Fierz transformation offers 
useful insight into the non-suppression of the $2\rarr 3$ process.

\subsection{Fierz Transformations in the Chiral Basis}
\label{subsec:FierzIntro}
Helicity projection operators are essential in chiral gauge
theories, so it is worth considering the reformulation of Fierz
transformations in the chiral basis~\cite{Nishi:2004st}.  (A
discussion of standard Fierz transformations may be found in,
e.g. Ref.~\cite{IZp161-2}.)  
We place hats above the generalized Dirac matrices constituting the chiral basis.
These matrices are
\bea{chiralbasis}
\{\Ghat^B\}&=&\{P_R,\,P_L,\,P_R\gamma^\mu,\,P_L\gamma^\mu,\,\half\sigma^{\mu\nu} \}\,,
\quad {\rm and} \quad \nonumber\\
\{\Ghat_B\}&=&\{P_R,\,P_L,\,P_L\gamma_\mu,\,P_R\gamma_\mu,\,\half\,\sigma_{\mu\nu} \}\,,
\eea
where $P_R\equiv\half (1+\gamma_5)$ and $P_L\equiv\half (1-\gamma_5)$ 
are the usual helicity projectors.
Notice that the dual of $P_R\gamma^\mu$ 
is $P_L\gamma_\mu$, and the dual of $P_L\gamma^\mu$ is $P_R\gamma_\mu$.
The tensor matrices in this basis contain factors of $\half$:
$\Ghat^T=\half\sigma^{\mu\nu}$ and $\Ghat_T = \half\sigma_{\mu\nu}$.
These facts result from the orthogonality and normalization properties
of the chiral basis and its dual, as explained in detail in Appendix~\ref{app:Fierz}.

Using completeness of the basis (see Appendix~\ref{app:Fierz}), one
arrives at a master formula which expands the outer product of two
chiral matrices in terms of their Fierzed forms:
\beq{chiralFierz1}
(\Ghat^D )\,[\Ghat_E ]=\frac{1}{4}\,Tr\,[\Ghat^D\,\Ghat^C\,\Ghat_E\,\Ghat_B ]\ (\Ghat^B ]\;[\Ghat_C )\,,
\eeq
where the parentheses symbols are a convenient shorthand for matrix
indices~\cite{Taka1986} (see the appendix for details).  Evaluating
the trace in Eq.~\rf{chiralFierz1} leads to the Fierz transformation
matrix in the chiral-basis:
\begin{widetext}
\bea{chiralFierz}
\left(
\barr{c}
(P_R)\ [P_R] \\
(P_L)\ [P_L] \\
(\hT)\ [\hT] \\
\!\!(\gamma_5\,\hT)\ [\hT] \\
(P_R)\ [P_L] \\
(P_R\gamma^\mu)\ [P_L\gamma_\mu] \\
(P_L)\ [P_R] \\
(P_L\gamma^\mu)\ [P_R\gamma_\mu] \\
(P_R\gamma^\mu)\ [P_R\gamma_\mu] \\
(P_L\gamma^\mu)\ [P_L\gamma_\mu] \\
\earr
\right) 
= \quarter
\left(
\barr{rrrr|rr|rr|r|r}
 2  & 0 & 1 & 1 &   &   &   &   &   &   \\ 
 0  & 2 & 1 &-1 &   &   &   &   &   &   \\
 6  & 6 &-2 & 0 &   &   &   &   &   &   \\
 6  &-6 & 0 & 2 &   &   &   &   &   &   \\ \hline
    &   &   &   & 0 & 2 &   &   &   &   \\
    &   &   &   & 8 & 0 &   &   &   &   \\ \hline
    &   &   &   &   &   & 0 & 2 &   &   \\
    &   &   &   &   &   & 8 & 0 &   &   \\ \hline
    &   &   &   &   &   &   &   &-4 & 0 \\ \hline
    &   &   &   &   &   &   &   & 0 &-4    
\earr
\right)
\ \left(
\barr{c}
(P_R]\ [P_R) \\
(P_L]\ [P_L) \\
(\hT]\ [\hT) \\
\!\!(\gamma_5\,\hT]\ [\hT) \\
(P_R]\ [P_L) \\
(P_R\gamma^\mu]\ [P_L\gamma_\mu) \\
(P_L]\ [P_R) \\
(P_L\gamma^\mu]\ [P_R\gamma_\mu) \\
(P_R\gamma^\mu]\ [P_R\gamma_\mu) \\
(P_L\gamma^\mu]\ [P_L\gamma_\mu) \\
\earr
\right)\,.
\eea
\end{widetext}
Non-explicit matrix elements in \rf{chiralFierz} are zero,
and we have introduced a shorthand $\hT$ for 
either $\Ghat^T=\half\sigma^{\mu\nu}$ or  $\Ghat_T=\half\sigma_{\mu\nu}$.

The importance of this transformation for us is that it converts
$t$-channel and $u$-channel exchange graphs into $s$-channel form, for
which it is straightforward to evaluate the partial waves.  The
block-diagonal structures, delineated with horizontal and vertical
lines, show that ``mixing'' occurs only within the subsets 
$\{ 
P_R\otimes P_R,\ P_L\otimes P_L,\ \hT\otimes \hT,\ \gamma_5\,\hT\otimes
\hT\}$, and $\{P_R\otimes P_L,\ P_R\gamma^\mu\otimes P_L\gamma_\mu 
\}$.
The Fierz transform matrix is idempotent, meaning its square is equal
to the identity matrix.  This follows from the fact that two Fierz
rearrangements return the process to its initial ordering.  A
consequence of the block-diagonal form is that each sub-block is
itself idempotent.

In Eq.~\rf{chiralFierz} we have included one non-member of the basis set,
namely $\gamma_5\,\hT$; it is connected to $\hT$ via the relation  
\beq{pT}
\gamma_5\,\sigma^{\mu\nu} = \frac{i}{2}\epsilon^{\mu\nu\alpha\beta}\sigma_{\alpha\beta}\,.
\eeq
Explicit use of $\gamma_5\,\hT$ in Eq.~\rf{chiralFierz} is an efficient way to express
the chiral Fierz transformation.

So far we have not used the qualifier in the assumption, that the dark
matter is Majorana.  Majorana particles are invariants under charge
conjugation $C$, which implies that vector and tensor bilinears are
disallowed.  Another way of understanding this is to note that
interchanging the two identical Majorana particles in a $t$-channel
diagram generates an accompanying $u$-channel diagram with a relative
minus sign (from fermion anticommutation).  When Fierzed, these two
amplitudes cancel for $V$ and $T$ couplings (exactly so in the
Four-Fermi limit where the differing momenta in the $t$- and
$u$-channel propagators can be ignored -- refer to Appendix
\ref{app:VectorTensor} for details).  We must thus drop $V$ and $T$
couplings appearing in the Fierzed bilinears of the $\chi$-current.

\subsection{Origin of $v^2$ and Helicity Suppressions}
\label{subsec:suppressions}

One can use partial wave decomposition (see. e.g., the
textbooks~\cite{Peskin,Weinberg,Srednicki}, or the convenient summary
in the Appendix of~\cite{BDJW}) to expand the scattering amplitudes as
a sum of angular momentum components.  Partial waves do not interfere,
and the $L^{th}$ partial wave contribution to the total cross section
$ \sigma v $ is proportional to $v^{2L}$.  
The annihilating $\chi$ particles are very non-relativistic today, so
an unsuppressed s-wave ($L=0$), if present, will dominate the
annihilation cross section.  The DM virial velocity within our Galaxy
is about $10^{-3}$ (in units of $c$), 
leading to a suppression of $v^2\sim 10^{-6}$ for
$p$-wave processes.

On the other hand, the SM fermions produced in the $2\rightarrow 2$
annihilation are highly relativistic (except possibly for $t\bar t$
production).  For many annihilation channels the spin state of the
fermion pair gives rise to a helicity suppression by a factor of
$(m_l/M_\chi)^2$, where $m_l$ is the fermion mass.

Unfortunately, many popular models for annihilation of Majorana dark
matter to charged leptons are subject to one or more of these two
suppressions, the $v^2$ and/or $(m_\ell/M_\chi)^2$ suppressions.  This
includes some of the models proposed to accommodate the positron and
$e^+ e^-$ excesses observed in PAMELA, Fermi-LAT, and HESS data.
In Section~\ref{sec:unsuppression}, we show that in the class of
models which have suppressed rates for $\chi\chi\rarr \ell^+\ell^-$,
the $2\rarr3$ graph obtained by adding a radiative $W^\pm$ or $Z$ to
the final state particles of the $2\rarr 2$ graph becomes
dominant. The radiated $W$'s and $Z$'s will decay to, among
other particles, antiprotons. Since an excess generation of antiprotons
is not observed by PAMELA, this class of models is ruled out by the
present work.

\begin{table*}[thdp]
\begin{center}
\begin{tabular}{||c|c||c|c||c|c||}
\hline
\multicolumn{2}{||c||}{s-channel bilinear $\Psibar\,\Gamma_D\,\Psi$} & \multicolumn{2}{c||}{$v=0$ 
limit} 
   & \multicolumn{2}{c||}{$M=0$ limit} \\ \cline{3-6}
\multicolumn{2}{||c||}{} & parallel spinors & antiparallel spinors & parallel spinors & antiparallel 
spinors \\  \hline\hline
scalar        & $\Psibar\,\Psi$                         
   & 0 & 0 & $ \sqrt{s}$ & 0 \\ \hline\hline
pseudo-scalar & $\Psibar\,i\gamma_5\,\Psi$               
   & $-2iM$ & 0 & $-i\sqrt{s}$ & 0   \\ \hline\hline
axial-vector  & $\Psibar\,\gamma_5\,\gamma^0\,\Psi$     
   & $2M$ & 0 & 0 & 0   \\ \cline{2-6}
              & $\Psibar\,\gamma_5\,\gamma^j\,\Psi$                                     
   & 0 & 0 & 0 & $\sqrt{s}\,(\pm\delta_{j1}-i\delta_{j2})$ \\ \hline\hline
vector        & $\Psibar\,\gamma^0\,\Psi$               
   & 0 & 0 & 0 & 0 \\ \cline{2-6}
              & $\Psibar\,\gamma^j\,\Psi$          
   & $\mp 2M\,\delta_{j3}$ & $-2M\,(\delta_{j1}\mp i\delta_{j2})$ & 0 & $-\sqrt{s}\,(\delta_{j1}\mp i
\delta_{j2})$ \\ \hline\hline
tensor        & $\Psibar\,\sigma^{0j}\,\Psi$        
   & $\mp 2iM\,\delta_{j3}$ & $-2iM\,(\delta_{j1}\pm\delta_{j2})$ & $-i\sqrt{s}\,\delta_{j3}$ & 0 \\ \cline
{2-6}
        & $\Psibar\,\sigma^{jk}\,\Psi$        
   & 0 & 0 & $\pm\sqrt{s}\,\delta_{j1}\delta_{k2}$ & 0 \\ \hline\hline
pseudo-tensor  & $\Psibar\,\gamma_5\,\sigma^{0j}\,\Psi$ 
   & 0 & 0 & $\pm i\sqrt{s}\,\delta_{j3}$ & 0 \\ \cline{2-6}
   & $\Psibar\,\gamma_5\,\sigma^{jk}\,\Psi$ 
   & $\mp 2M\,\delta_{j1}\delta_{k2}$ & $-2M\,(\delta_{j2}\delta_{k3}\mp i\delta_{j3}\delta_{k1})$ 
         & $-\sqrt{s}\,\delta_{j1}\delta_{k2}$ & 0 \\ \hline\hline
\end{tabular}
\end{center}
\caption{
Extreme non-relativistic and extreme relativistic limits for s-channel bilinears.
In order for a term with an initial-state DM bilinear and a final-state lepton bilinear to remain unsuppressed,
the DM bilinear must have a non-zero entry in the appropriate cell of the ``$v=0$ limit'' columns,
and the lepton bilinear must have a non-zero term in the appropriate cell of the  ``$M=0$ limit'' columns. 
Otherwise, the term is suppressed. 
(The tensor and pseudo-tensor are not independent, but rather are related by 
$\gamma_5\,\sigma^{\mu\nu}=\frac{i}{2}\epsilon^{\mu\nu\alpha\beta}\,\sigma_{\alpha\beta}$.)
We recall that antiparallel spinors correspond to parallel particle spins 
(and antiparallel particle helicities for the $M=0$ current), and vice versa.
Amplitudes are shown for $\ubar\,\Gamma_D\,v = [ \vbar\,\Gamma_D\,u ]^*$.
The two-fold $\pm$ ambiguities reflect the two-fold spin assignments for parallel spins, and 
separately for antiparallel spins.
}
\label{table:bilinearlimits}
\end{table*}

Consider products of s-channel bilinears of the form $(\chibar \;
\Gamma_1 \chi)(\bar l \; \Gamma_2 l)$.
To further address the question of which products of currents are
suppressed and which are not, we may set $v^2$ to zero in the
$\chi$-current, and $m_\ell^2$ to zero in the lepton current, and ask
whether the product of currents is suppressed.  If the product of
currents is non-zero in this limit, the corresponding amplitude is
unsuppressed.  In Table \ref{table:bilinearlimits} we give the results
for the product of all standard Dirac bilinears.  (The derivation of
these results is outlined in Appendix \ref{app:NRnERlimits}.)
Suppressed bilinears enter this table as zeroes.
\footnote{It is seen that the only bilinears in the table without velocity-suppression are those of the pseudo-scalar, the three-vector part of the vector, the zero$^{\rm th}$ component of the axial vector, and the time-space part of the tensor (or equivalently, the space-space part of the pseudotensor).  It is also seen that the only bilinears without fermion mass-suppression are the scalar, pseudoscalar, three-vector parts of the vector and axial vector, and the tensor.}

One can read across rows of this table to discover that the only
unsuppressed $s$-channel products of bilinears for the $2\rarr 2$
process are those of the pseudo-scalar, vector, and tensor.  (For
completeness, we also show results for the pseudo-tensor bilinears,
although the pseudo-tensor is not independent of the tensor, as a
result of Eq.~\ref{pT}.)  For Majorana dark matter, the vector and
tensor bilinears are disallowed by charge-conjugation arguments 
and one is left with just the unsuppressed pseudo-scalar.

\subsection{Class of Models for which $\chi\chi\rarr\ell\ellbar$ Annihilation is Suppressed }
\label{subsec:suppression}

We now put the results of the previous two subsections together to
explain which class of models have a $v^2$ and/or $(m_\ell/M_\chi)^2$
suppressed $2\rarr 2$ annihilation.  
We have seen that, for Majorana DM, $s$-channel annihilation with a
$P$ coupling is unsuppressed, while $S$ and $A$ contributions
are suppressed (and $V$ and $T$ forbidden).  Let us now consider
$t$-channel or $u$-channel processes.

Any $t$-channel or $u$-channel diagram that Fierz's to an $s$-channel
form containing a pseudoscalar coupling will have an unsuppressed
$L=0$ $s$-wave amplitude.  From the matrix in Eq.~(\ref{chiralFierz}),
one deduces that such will be the case for any $t$- or $u$-channel
current product on the left side which finds a contribution in the
$1^{st}$, $2^{nd}$, $5^{th}$, or $7^{th}$ columns of the right side.
This constitutes the $t$- or $u$-channel tensor, same-chirality
scalar, and opposite chirality vector products (rows 1 through 4, and
6 and 8 on the left).
On the other hand, the $t$- or $u$-channel opposite chirality scalars
or same-chirality vectors (rows 5, 7, 9, and 10 on the left) do {\it 
not} contain a pseudoscalar coupling after Fierzing to $s$-channel form.  
Rather, it is the suppressed axial-vector and vector (Dirac fermions only) that appears.

Interestingly, a class of the most popular models for fermionic dark
matter annihilation to charged leptons, fall into this latter,
suppressed, category. It is precisely the opposite-chirality $t$- or
$u$-channel scalar exchange that appears in these models, an explicit
example of which will be discussed below.
Thus it is rows 5 and 7 in Eq.~(\ref{chiralFierz}) that categorize the
model we will analyze.  After Fierzing to $s$-channel form, it is seen
that the Dirac bilinears are opposite-chirality vectors (i.e., $V$ or
$A$).  Dropping the vector term from the $\chi$-current we see that
the $2\rightarrow 2$ process couples an axial vector $\chi$-current to
a relativistic SM fermion-current which is an equal mixture
of $A$ and~$V$.
Accordingly, this model has an $s$-wave amplitude occurring
only in the $L=0$, $J=1$, $S=1$ channel, with the spin flip from $S=0$
to $S=1$ (or equivalently, the mismatch between zero net chirality and
one unit of helicity) costing a fermion mass-insertion and a
$(m_f/M_\chi)^2$ suppression in the rate.

Let us pause to explain why this $t$- or $u$-channel scalar exchange
with opposite fermion chiralities at the vertices is so common.  It
follows from a single popular assumption, namely that the dark matter
is a gauge-singlet Majorana fermion.  As a consequence of this
assumption, annihilation to SM fermions, which are $SU(2)$ doublets or singlets,
requires either an $s$-channel singlet boson or a $t$- or $u$-channel
singlet or doublet scalar that couples to $\chi$-$f$.  In the first
instance, there is no symmetry to forbid a new force between SM
fermions, a disfavored possibility.  In the second instance, unitarity
fixes the second vertex as the hermitian adjoint of the first.  Since
the fermions of the SM are left-chiral doublets and right-chiral
singlets, one gets chiral-opposites for the two vertices of the $t$-
or $u$-channel.

Supersymmetry provides an analog of such a model.  In this case the
dark matter consists of 
Majorana neutralinos, which
annihilate to SM fermions via the exchange of (``right''- and
``left''-handed) $SU(2)$-doublet slepton fields.  
In fact, the implementation in 1983 of supersymmetric photinos as dark matter 
provided the first explicit calculation of $s$-wave suppressed 
Majorana dark matter~\cite{Haim1983}.
However, the class of models described above is more general than the 
class of supersymmetric models.

To illustrate our arguments, we choose a simple example of the class
of model under discussion.  This is provided by the leptophilic
model proposed in Ref.~\cite{Cao:2009yy} by Cao, Ma and Shaughnessy.
Here the DM consists of a gauge-singlet Majorana fermion $\chi$ which
annihilates to leptons via the $SU(2)$-invariant interaction term
\begin{equation}
f\left(\nu\,\ell^-\right)_L\,\varepsilon\,
\left(
\barr{l}
\eta^+ \\
\eta^0 \\
\earr
\right)\chi + h.c.
= f(\nu_L\eta^0 - \ell_L^{-} \eta^+)\chi + h.c.
\label{eq:ma}
\end{equation}
where $f$ is a coupling constant, $\varepsilon$ is the
$2\times 2$ antisymmetric matrix, and $(\eta^+$, $\eta^0)$ form
the new $SU(2)$ doublet scalar which mediates the annihilation.  (This
model was originally discussed in Ref.~\cite{Ma:2000cc}, and an expanded 
discussion of its cosmology may be found in Ref.~\cite{Gelmini}.)

As discussed above, the $u$- and $t$-channel amplitudes for DM
annihilation to leptons, of the form $(\chibar P_L l)\,(\bar l P_R
\chi)$, become pure $(\chibar P_L\gamma^\mu \chi)\,(\bar l
P_R\gamma_\mu l)$ under the chiral Fierz transformation.
The product of the Majorana and fermion bilinears then leads to an
$AA$ term and an $AV$ term. However, reference to
Table~\ref{table:bilinearlimits} shows that neither of these terms
leads to an unsuppressed amplitude: in all cases, either the lepton
bilinear is suppressed by $m_\ell$, the DM bilinear by $v$, or both
are suppressed.  Thus, Majorana DM annihilation to a lepton
pair is suppressed in this model, in accordance with the explicit
calculation in Ref.~\cite{Cao:2009yy}.

\section{Lifting the Suppression}
\label{sec:unsuppression}

\begin{figure*}[ht]
\includegraphics[width=14cm]{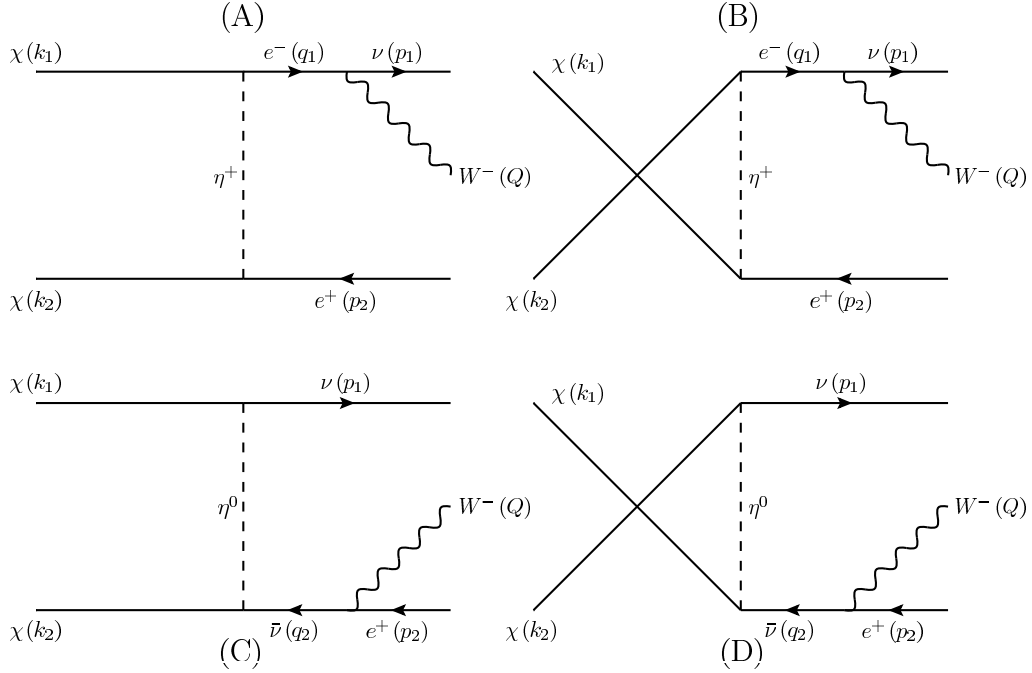}
\caption{
$t$-channel (A and C) and $u$-channel (B and~D) contributions 
  to $\chi\chi\rightarrow e^+ \nu W^-$. Emission
  from the scalar propagator is not included, as it is suppressed by
  $1/M_\eta^2$. Note that all fermion momenta flow with the arrow
  except $p_2$, so $q_1=p_1+Q$, $q_2=-p_2-Q$.
\label{feyngraph1}}
\end{figure*} 

Allowing the lepton bilinear to radiate a $W$ or $Z$ boson (as shown in Fig.~\rf{feyngraph1}) 
does yield an unsuppressed amplitude.
In the rate, there will be the usual radiative suppression factor of $\frac{\alpha_2}{4\pi}\sim 10^{-3}$. 
But, this will be partially compensated by a 3-body phase space factor $\sim (M_\chi/M_W)^2/8\pi^2$ relative 
to 2-body massless phase space, which exceeds unity for dark matter masses exceeding~$\sim$TeV\@.\footnote{When $M^2_\chi \gg M^2_W$, the rate for single W production is dominated by infrared and collinear divergences, leading to a suppressed factor $\ln^2 \left( \frac{s}{4M^2_W} \right) - 3 \ln \left( \frac{s}{4M^2_W} \right)$~\cite{Ciafaloni:2010qr,Berezinsky} instead of our $\left( \frac{s}{4M^2_W} \right)$. Moreover, the rate for multiple production of $W$'s becomes so large that resummation techniques are necessary.}
More importantly, the $v^2$ suppression for Majorana annihilation to 2-body final states will be lifted
by the 3-body $W$-bremsstrahlung process.
In Section \ref{sec:example} we show, by explicit calculation, that the $2\rarr 3$ radiative process that 
leads to antiprotons 
dominates for any $M_\chi$ that allows the $W$ to be produced on-shell, 
i.e., for $2M_\chi> M_W$. 

The next inevitable question is ``Why is the radiative $2\rarr 3$ process unsuppressed?''
To answer this question, we invoke a more general Fierz rearrangement applicable to $2\rarr N$ 
processes, $N\ge 3$.
The relevant equation, derived in Appendix~\ref{app:Fierz} states that 
any $4\times 4$ matrices $\X$ and $\Y$ may be expressed as 
\bea{genFierz}
(\X)\,[\Y] 
&& = (\X\openone)\ [\openone \Y]  \nonumber \\
&& = \frac{1}{4}\,(\X\,\Gamma^B\,\Y\,]\ [\,\Gamma_B\,) \nonumber \\
&& =  \frac{1}{4^2}\,Tr\,[\X\,\Gamma^B\,\Y\,\Gamma_C ]\ (\Gamma^C\,]\ [\Gamma_B\,)\,,
\eea
where the Dirac matrices here are taken in the 
standard basis defined in Eq.~(\ref{stdbasis}).

From Table~\rf{table:bilinearlimits} we see that setting $\Gamma^C$ to $\gamma_5\,\gamma^0$,
the only structure available to a non-relativistic Majorana current other than the pseudoscalar,
and 
$\Gamma_B$ to either $\gamma^j$ or $\gamma_5\gamma^j$, 
provides an unsuppressed product of the Majorana dark matter bilinear and the 
charged lepton bilinear.
Moreover, for the $W/Z$-bremsstrahlung process, 
$X$ and $Y$ in the general Fierz equation are the un-Fierzed couplings 
$P_L$ and $q^{-2}\,P_R\,\slashed{q}\,P_L\,\slashed{\epsilon}$, respectively.
So we will have shown that the radiative process is unsuppressed if we can show that 
$q^{-2}\,Tr\,[P_L\,(\gamma^j{\rm\ or\ }\gamma_5\gamma^j)\,P_R\,\slashed{q}\,P_L\,\slashed{\epsilon}\,\gamma_5\,\gamma_0 ]$ 
is unsuppressed.
This trace reduces to 
$q^{-2}\,Tr\,[P_R\,\gamma_0\,\gamma^j\,\slashed{q}\,\slashed{\epsilon} ]$. 
The expansion of this trace as scalar products contains terms such as $q_0\cdot \epsilon^j$
and $({\vec\epsilon}\times{\vec q})^j$, which are nonzero and unsuppressed by fermion masses.
Thus, the $2\rarr 3$~process contains an unsuppressed $s$-wave amplitude.

Physically, the un-suppression works because the gauge boson carries
away a unit of angular momentum,
allowing a fermion spin-flip such that there is no longer a mismatch between
the chirality of the leptons and their allowed two-particle spin state.

One may ask why emission of a gamma-ray rather than a $W/Z$~boson is less effectual.
It has been known for some time~\cite{hardgamma,Bringmann:2007nk} 
that gamma-ray emission in the final state 
does produce an unsuppressed $s$-wave contribution, but at second order 
rather than lowest order in the inverse mass-squared $M_\eta^{-2}$ of the $t$- and $u$-channel exchange particle(s).
The reason is that gamma-ray emission from the final state fermions~(FSR) 
is dominated by infra-red and collinear singularities,
each of which puts the intermediate lepton on-shell (virtuality~$q^2\rarr 0$).
Including the $q^{-4}$ from the squared propagator in the phase space integral (see Eq.~\rf{phasespace}),
one gets the factor $\int_{M_V^2}^s \frac{dq^2}{q^6}(s-q^2)\,(q^2-M_V^2)$,
where $M_V$ is the mass of the radiated boson~ (photon or $W$ or~$Z$). 
For a gamma-ray, with $M_V^2=0$, one readily sees the infra-red and collinear singularities 
in $\int_0 \frac{dq^2}{q^4}$.
An on-shell particle is observable, so the spin states of the $q^2\rarr 0$ intermediate fermion do not interfere.
Thus, as $q^2\rarr 0$, the trace for gamma emission, 
$Tr\,[\gamma_0\,\gamma^j\,\slashed{q}\,\slashed{\epsilon} ]
 = Tr\,[\gamma_0\,\gamma^j\,(P_R^2+P_L^2)\slashed{q}\,\slashed{\epsilon} ]$
goes over to 
$Tr\,[\gamma_0\,\gamma^j\,P_R]\,Tr\,[P_R\slashed{q}\,\slashed{\epsilon} ]
+ Tr\,[\gamma_0\,\gamma^j\,P_L]\,Tr\,[P_L\slashed{q}\,\slashed{\epsilon} ]$.
The first trace in each term of this sum vanishes.
Consequently, the gamma-emission amplitude remains suppressed at order~$M_\eta^{-2}$.
However, at order~$M_\eta^{-4}$, the gamma-ray may be emitted from the 
internal particle $\eta$
~(VIB).
For VIB, phase space does not favor $q^2=0$, and an unsuppressed amplitude results.
 
The emission of a massive $W$ (or~$Z$) boson contrasts significantly from 
the emission of a massless photon.
With the $W$ emission, the relevant phase space integral over virtuality~$q^2$ is
$\int_{M_W^2}^s \frac{dq^2}{q^6}(s-q^2)\,(q^2-M_W^2)$.
The minimum virtuality of the intermediate fermion is 
$q^2 = M_W^2$, and the mean virtuality for $s\gg M_W^2$ is greater again by the factor 
$2\ln(s/M_W^2)$. 
With no infra-red or collinear singularities for $W/Z$-emission,
an unsuppressed amplitude results already at order~$M_\eta^{-2}$. 

Before looking at an explicit example in which electroweak
bremsstrahlung is seen to lift a suppression, we pause to summarize
some important facts
for the $2\rarr 2$ annihilation process:

\begin{itemize}

\item
Fierz transformation is used to re-express $t$- and $u$-channel amplitudes
of the form $(\chibar \; \Ghat^A l)(\bar l \; \Ghat_B \chi)$ as a
sum of $s$-channel (not to be confused with $s$-wave) amplitudes 
of the form $(\chibar \; \Ghat^C \chi)(\bar l \; \Ghat_D l)$.

\item
For Majorana dark matter, only $S$, $P$, and $A$ $s$-channel bilinears are
allowed, with the $V$ and $T$ bilinears forbidden by the self-conjugate
properties of Majorana particles.

\item
Considering the product of an $s$-channel $\chi$-current with an
$s$-channel fermion-current, we find that the 
pseudo-scalar is the only 
member of the set ($S$, $P$, $A$) which is unsuppressed.  The other
combinations are either helicity $(m_\ell/m_\chi)$ or velocity~$(v)$
suppressed.

\item
The annihilation process $\chi\chi\rarr\ell\ellbar$ via $t$- and
$u$-channel exchange of a scalar is suppressed. 
Importantly, electroweak bremsstrahlung lifts this suppression 
at lowest order in the propagator mass-squared ($M_\eta^{-2}$ in amplitude), 
whereas photon bremsstrahlung lifts the suppression at the next order 
($M_\eta^{-4}$ in amplitude).

\end{itemize}

Amplification of the latter remark is the purpose of this paper.

\section{Explicit Calculation of Suppression-lifting with Electroweak Bremsstrahlung}
\label{sec:example}

To explicitly demonstrate that emission of a $W^\pm$ or $Z$ boson does lift
helicity suppression, we calculate the cross section for
$\chi\chi\rightarrow\e^\mp\overset{(-)}{\nu} W^\pm$ below in the leptophilic
model of Ref~\cite{Cao:2009yy}.
The interaction term for this model is that given above in Eq.~(\ref{eq:ma}).

\subsection{Example of Helicity-Suppressed Rate}
\label{subsec:Ma}
In the model of ref.~\cite{Cao:2009yy},
the cross section for the $2\rarr 2$ process 
$\chi\chi\rightarrow\e^+e^-$~or~$\nu\nubar$ 
with Majorana DM is given as
\begin{eqnarray}
v\,\sigma = \frac{f^4 v^2 r^2}{24\pi\,M_\chi^2}\,(1-2r+2r^2)\,,
\label{eq:tree}
\end{eqnarray}
where $m_l\simeq0$ and $M_{\eta^\pm}=M_{\eta^0}$ have been assumed,
and $r=M_\chi ^2/(M_\eta ^2+M_\chi ^2)$.
The suppressions discussed in Section \ref{sec:Suppression} are
apparent in Eq.~(\ref{eq:tree}).  The helicity suppressed $s$-wave term
is absent in the $m_l\ =0$ limit, and thus only the 
$v^2$-suppressed term remains.

This $2\rarr 2$ cross-section can be calculated by inclusion of two 
Feynman diagrams, a $t$-channel exchange of $\eta$ and the associated $u$-channel exchange obtained by crossing the Majorana particles.
The relative sign between the graphs is negative, due to the fermion exchange.
Summing and squaring, one has three terms including the interference term.
Alternatively, one may Fierz transform the fermion bilinears in the two contributing amplitudes.
The relative minus sign is compensated by the special Majorana minus sign described in 
Eq.~\rf{Majswap}.  Reference to Eq.~\rf{chiralFierz} then shows that one gets 
$(P_L)\,[P_R]\rarr \half\,(P_L\gamma^\mu]\,[P_R\gamma_\mu)\times 2$,
where the final factor of 2 counts the two contributing amplitudes,
which are identical in the four-fermi limit $M_\eta^2\gg t{\rm \ and\ }u$.
We are left with just one amplitude,  
$\frac{f^2}{M_\eta^2}[\vbar(k_2)(\half\gamma_5) v(p_2)]\,[\ubar(p_1)P_L \gamma_\mu v(p_2)]$.
The surviving Dirac structure for the Majorana current is pure axial vector,
since the vector (and tensor) part of a Majorana current vanishes.
With just a single product of bilinears, 
the remaining part of the $2\rarr 2$ calculation is straightforward.
One arrives at
\be
\label{Fierzed2to2}
v\,\sigma = \frac{f^4\,M_\chi^2}{16\pi\,M_\eta^4}
\left[ \frac{m_l^2}{s} +\frac{2}{3}v^2 +{\cal O}(v^4)\right]\,,
\ee
in agreement with the four-fermi, $m_\ell=0$ limit of 
Eq.~\rf{eq:tree}. 
Here, the helicity suppression of the $s$-wave amplitude, proportional to a helicity flip, 
in turn proportional to a mass insertion, is manifest.

\subsection{{\bm $W$} Emission and Unsuppressed {\bm $S$}-wave}
\label{subsec:Wemmission}

We now turn to the calculation of the cross section for the process
$\chi\chi\rightarrow e^+\nu W^-$ (equal to that for
$\chi\chi\rightarrow e^-\bar\nu W^+$).  The four contributing Feynman
diagrams are shown in Fig.~\ref{feyngraph1}.  Note that we consider
bremsstrahlung only from the final state particles (FSR), and neglect emission from the
virtual scalar (VIB).  Strictly speaking, the distinction between FSR
and VIB is somewhat artificial in the sense that the partition depends
upon the choice of gauge.  However, we shall work in unitary gauge, in
which emission from the internal line is suppressed by a further power
of $M_\eta^2$ due to the additional scalar propagator; consequently,
we expect our results to be valid to order $M_\eta^{-2}$ in amplitude,
i.e. order $M_\eta^{-4}$ in rate.
  
We retain the assumptions $m_l\simeq 0$ and $M_{\eta^\pm}=M_{\eta^0}$.
The matrix element for the top-left diagram is
\begin{eqnarray}
\mathcal{M}_{A} &=&\frac{ig f^2}{\sqrt{2}q_1^2}\frac{1}{{t_1}-M_\eta^2}\Big(\bar v(k_2) P_Lv
(p_2)\Big) \nonumber\\
&&\times \Big(\bar u(p_1) \gamma^\mu P_L \slashed{q_1} u(k_1)\Big)\epsilon^Q_\mu.
\end{eqnarray}
where $i\frac{g}{\sqrt{2}}\gamma^\mu P_L$ is the coupling at the $\ell\nu W$ vertex, and $t_1,t_2,u_1,u_2$ are the standard Mandelstam variables,
\begin{eqnarray}
t_1&=(k_{1}-q_{1})^2=&(p_{2}-k_{2})^{2}\nonumber\\ 
t_2&=(k_{1}-p_{1})^2=&(-q_{2}-k_{2})^{2}\nonumber\\
u_1&=(k_2-q_1)^2=&(p_2-k_1)^2\nonumber\\
u_2&=(k_2-p_1)^2=&(-q_2-k_1)^2.
\end{eqnarray}
Upon applying Eq~(\ref{genFierz}) to Fierz transform the matrix element, 
we obtain
\begin{eqnarray}\label{MA}
\mathcal{M}_{A}&=&\frac{ig f^2}{\sqrt{2}q_1^2}\frac{1}{{t_1}-M_\eta^2}\epsilon^Q_\mu\frac{1}
{4}\nonumber\\
&&\Big[\Big(\bar v(k_2) u(k_1)\Big)\Big(\bar u(p_1 )P_L \gamma^\mu P_L\slashed{q_1} v(p_2)\Big)
\nonumber\\
&&+\Big(\bar v(k_2) \gamma_5 u(k_1)\Big)\Big(\bar u(p_1 )P_L \gamma_5\gamma^\mu P_L
\slashed{q_1} v(p_2)\Big) \nonumber\\
&&+\Big(\bar v(k_2) \gamma_5 \gamma_\alpha u(k_1)\Big)\Big(\bar u(p_1 )\gamma^\alpha 
\gamma^\mu P_L\slashed{q_1} v(p_2)\Big)\Big]\nonumber\\
&=&\frac{ig f^2}{\sqrt{2}q_1^2}\frac{1}{{t_1}-M_\eta^2}\epsilon^Q_\mu\frac{1}{4}\nonumber\\
&\times&\left(\bar v(k_2) 
\gamma_5 \gamma_\alpha u(k_1)\Big)\Big(\bar u(p_1 ) P_L\gamma^\alpha \gamma^\mu\slashed
{q_1} v(p_2)\right).
\end{eqnarray} 
The first two terms after the first equality are zero due to the
helicity projection operators, leaving only an axial vector term.
(Vector and tensor $\chi$-bilinears have been omitted, as they will
cancel between $u$ and $t$ channel diagrams in the heavy $M_\eta$
limit, as discussed above.)
Note that although this matrix element resembles that of an $s$-channel
annihilation process, the $\gamma$ matrices in the lepton bilinear would
be in a different order for a true $s$-channel annihilation process
involving $W/Z$-bremsstrahlung from one of the final state leptons.

Similarly, the matrix element for the top-right diagram can be written as
\begin{eqnarray}\label{MB}
\mathcal{M}_{B} &=&\frac{-ig f^2}{\sqrt{2}q_{1}^2}\frac{1}{u_1-M_\eta^2}\frac{1}{4}\Big(\bar v
(k_2)\gamma_5 \gamma_\alpha u(k_1)\Big)\nonumber\\
&&\times\Big(\bar u(p_1 )P_L\gamma^\alpha \gamma^\mu 
\slashed{q_1} v(p_2)\Big)\epsilon^Q_\mu,
\end{eqnarray}
and those for the bottom diagrams,
\begin{eqnarray}
\mathcal{M}_{C}&=&\frac{-ig f^2}{\sqrt{2}q_2^2}\frac{1}{t_2-M_\eta^2}\frac{1}{4}\Big(\bar v
(k_2)\gamma_5 \gamma_\alpha u(k_1)\Big)\nonumber\\
\label{MC}&&\times\Big(\bar u(p_1 )P_L\slashed{q_2} \gamma^\mu
\gamma^\alpha  v(p_2)\Big)\epsilon^Q_\mu,\\
\mathcal{M}_{D} &=&\frac{ig f^2}{\sqrt{2}q_2^2}\frac{1}{u_2-M_\eta^2}\frac{1}{4}\Big(\bar v
(k_2)\gamma_5 \gamma_\alpha u(k_1)\Big)\nonumber\\
\label{MD}&&\times\Big(\bar u(p_1 )P_L\slashed{q_2} \gamma^\mu
\gamma^\alpha  v(p_2)\Big)\epsilon^Q_\mu.
\end{eqnarray}
Performing the sum over spins and polarizations, we find
\begin{widetext}
\begin{eqnarray}
&&\sum_\text{spin, pol.} |\mathcal{M}|^2 = \sum_\text{spin, pol.} |\left(\mathcal{M}_{A} +\mathcal{M}_{C}\right)-\left( \mathcal{M}_{B} +\mathcal{M}_{D}\right)|^2\nonumber\\
&=& \left( \frac{g f^2}{\sqrt{2}}\right)^2\frac{1}{16}{\rm Tr}\left[(\slashed{k_2}+M_\chi)\gamma_\alpha 
(\slashed{k_1}+M_\chi)\gamma_\beta\right]
\left(g_{\mu\nu} - \frac{Q_\mu Q_\nu}{M_W^2}\right)\Bigg(\frac{1}{q_{1}^{4}}\left(\frac{1}{t_1-M_
\eta^2} + \frac{1}{u_1-M_\eta^2}\right)^2{\rm Tr}\Big[ \slashed{p_1} \gamma^\alpha \gamma^\mu 
\slashed{q_{1}} \,\slashed{p_2} \,\slashed{q_{1}}\, \gamma^\nu \gamma^\beta P_R\Big]\nonumber\\
&&-\frac{1}{q_{1}^{2}q_{2}^{2}}\left(\frac{1}{t_1-M_\eta^2} + \frac{1}{u_{1}-M_\eta^2}\right)\left(\frac
{1}{t_2-M_\eta^2} + \frac{1}{u_2-M_\eta^2} \right)
\bigg({\rm Tr}\Big[ \slashed{p_1} \gamma^\alpha 
\gamma^\mu \slashed{q_{1}} \,\slashed{p_2} \, \gamma^\beta \gamma^\nu \slashed{q_{2}}\, P_R
\Big]+{\rm Tr}\Big[ \slashed{p_1}  \slashed{q_{2}}  \gamma^\mu \gamma^\alpha  \,\slashed{p_2} \,
\slashed{q_{1}}\, \gamma^\nu \gamma^\beta P_R\Big]\bigg)\nonumber\\
&&+\frac{1}{q_{2}^{4}}\left(\frac{1}{t_{2}-M_\eta^2} + \frac{1}{u_{2}-M_\eta^2}\right)^2{\rm Tr}\Big
[ \slashed{p_1} \slashed{q_{2}}  \gamma^\mu \gamma^\alpha  \,\slashed{p_2} \, \gamma^\beta 
\gamma^\nu \slashed{q_{2}}\, P_R\Big]\Bigg)\nonumber\\
\label{MatElement}
\end{eqnarray}
\end{widetext}
We evaluate this in terms of scalar products using the standard Dirac Algebra, 
leading to a result too lengthy to record here.

The thermally-averaged rate
is given by
\begin{equation}
v\,d\sigma = \frac{1}{2s} \int\frac{1}{4} \sum_\text{spin, pol.} |\mathcal{M}|^2 
 \, dLips^3
\label{dsigma1}
\end{equation}
where the $\quarter$ arises from averaging over the spins of the initial $\chi$ pair,
and $v=\sqrt{1-\frac{4M_\chi^2}{s}}$ is the mean dark matter relative velocity,
as well as the dark matter single-particle velocity in the center of mass frame\footnote
{Informative discussions of the meaning of $v$ are given in~\cite{LL}, 
and, including thermal averaging, in~\cite{GelmGondo}.}.

The three-body Lorentz Invariant Phase Space is
\begin{eqnarray}
&dLips^3 = (2 \pi)^4 
\dfrac{d^3 \vec p_1}{2E_1}\dfrac{d^3\vec p_2}{2E_2}\dfrac{d^3 \vec Q}{2E_W} \dfrac{\delta^4(P-
p_1-p_2-Q)}{(2\pi)^9}
\end{eqnarray}
and $P=k_1+k_2$. This factorizes into the product of two two-body phase space integrals, 
convolved with an integral over the fermion propagator momentum,
\begin{eqnarray}
&dLips^3=\int_{M_W^2}^s\dfrac{dq_1^2}{2\pi}\left(\dfrac{d^3\vec q_1}{2E_{q_1}}\dfrac{d^3 \vec p_2}{2E_2} \dfrac{\delta^4(P-q_1-p_2)}{(2\pi)^2 }\right)\nonumber\\
&\times \left(\dfrac{d^3 \vec p_1}{2E_1}\dfrac{d^3 \vec 
Q}{2E_W} \dfrac{\delta^4(q_1-Q-p_1)}{(2\pi)^2}\right)\nonumber\\
&=\int_{M_W^2}^s\dfrac{dq_1^2}{2\pi}\,dLips^2(P^2,q_1^2,p_2^2)\,dLips^2(q_1^2,Q^2,p_1^2).
\end{eqnarray}
Evaluating the two-body phase space factors in their respective center
of momentum frames, and using $p_1^2=p_2^2 = 0$, we have
\begin{eqnarray}
 dLips^2(x^2,y^2,0) = \dfrac{x^2-y^2}{8\pi x^2} \dfrac{d\bar\Omega}{4\pi}.
\end{eqnarray}
This allow us to write the three-body phase space as
\begin{eqnarray}
\label{phasespace}
dLips^3&=& \frac{1}{2^6 (2 \pi)^4} \int_{M_W^2}^s dq_1^2 \\
&\times&\frac{(s-q_1^2)(q_1^2-Q^2)}{s q_1^2} \, d\phi\, d \cos 
\theta_P \,d\cos\theta_q, \nonumber
\end{eqnarray}
where $\phi$ is the angle of intersection of the plane defined by
$\chi\chi \rightarrow ee^*$ with that defined by $e \nu W$, and $\theta_P$ and 
$\theta_q$ are defined in $P$ (CoM) and $q$ rest frames respectively.

\begin{figure}[t]
\includegraphics[width=1.0\columnwidth]{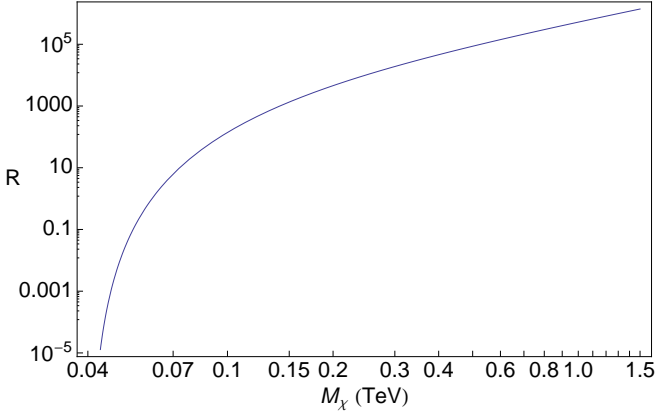}
\caption{
\label{ratio1} 
The ratio $R=v\,\sigma (\chi\chi\to e^+ \nu W^-) /v\,\sigma
(\chi\chi\to e^+e^-)$ for the example model~\cite{Cao:2009yy}, with
$M_\eta^2\gg M_\chi^2$.  We have used $v = 10^{-3}c$, appropriate for
the Galactic halo.}
\end{figure} 

\begin{figure}[t]
\includegraphics[width=1.0\columnwidth]{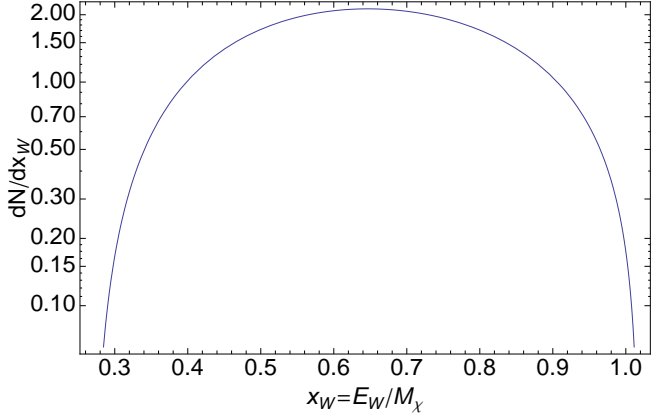}
\caption{
\label{dNdEW}
The $W$ spectrum per $\chi\chi\to e \nu W$ event for the example model, 
with $M_\chi=300$ GeV and $M_\eta^2\gg M_\chi^2$.  
}
\end{figure} 

We evaluate the scalar products that arise from Eq.(\ref{MatElement})
in terms of the invariants $q_1^2$, $Q^2=M_W^2$, $s$, $t_1$, and
$u_1$, and the angles $\theta_P$, $\theta_q$, and $\phi $. We then use
Eq.~(\ref{dsigma1}) to evaluate the cross section. As we have neglected
diagrams suppressed by $M_\eta^{-2}$ relative to those in
Fig.~\ref{feyngraph1}, we present our results to leading order in 
$M_\eta^{-4}$  
(i.e., we take $M_\eta^2 \gg t_1,t_2,u_1,u_2$).  
To leading order in powers of $M_\chi$ and $M_W$ in the numerator, we find 
\begin{align}
v\,\sigma  &=\frac{g^2 f^4}{512 M_W^2 M_\eta^4 \pi ^3}
\Bigg\{
M_\chi^4 
\left( \frac{1}{3}\ln\left[\frac{4 M_\chi^2}{M_W^2}\right] - \frac{7}{18} \)
\nonumber\\&
+ M_\chi^2  M_W^2 \Bigg(
\ln\left[\frac{4 M_\chi^2}{M_W^2}\right] 
\left\{ 1+\ln\left[\frac{2M_W M_\chi}{M_W^2+4 M_\chi^2}\right] \right\}
\nonumber\\&
-1 + \text{Li}_2\left[\frac{4 M_\chi^2}{M_W^2+4 M_\chi^2}\right]
- \text{Li}_2\left[\frac{M_W^2}{M_W^2+4 M_\chi^2}\right]
\Bigg) 
\nonumber\\&
+ \Order(M_W^4)
\Bigg\}.\label{vsigma}
\end{align}
The Spence function (or ``dilogarithm'') is defined as 
${\rm Li_2} (z)\equiv -\int^z_0 \frac{d\zeta}{\zeta}\ln|1-\zeta|
   = \sum_{k=1}^\infty \frac{z^k}{k^2}$.
The full expression (retaining sub-leading terms in $M_\chi$ in the numerator) is
specified in Appendix~\ref{app:FullResults}.  Clearly, the leading
terms are neither helicity nor velocity suppressed.

The effectiveness of the $W$-strahlung processes in lifting
suppression of the annihilation rate can be seen Fig.~\ref{ratio1},
where we plot the ratio of the $W$-strahlung cross section to that of
the lowest order process, $R_W = v\,\sigma (\chi\chi\to e^+ \nu W^-)
/v\,\sigma (\chi\chi\to e^+e^-)$.  
We see that the $W$-strahlung rate rises with DM mass, to quickly dominate over the lowest order annihilation process.
The $W$-bremsstrahlung rate rises approximately as $M^4_\chi$.
As $M^2_\chi$ increases, eventually
phase space allows multi-$W/Z$ radiative production, with such a large
rate that resummation techniques become necessary.  The onset of
multi-$W/Z$ dominance has been discussed in~\cite{Berezinsky:2002hq,Kachelriess,BDJW}.

To obtain the energy spectrum of the $W$, we compute the differential
cross section in terms of $E_W$ by making the transformation
\begin{eqnarray}
d\cos(\theta_q) \rightarrow \frac{-4 \sqrt{s} q^2}{(s-q^2)(q^2-M_W^2)} dE_W. 
\end{eqnarray}
We find~\cite{rapidity}, again to leading order $M_\eta^{-4}$,
\begin{eqnarray}
\label{dsigdEW}
 &&\frac{v\,d\sigma}{dE_W} =\frac{g^2 f^4}{512 E_W M_W^2 M_\eta^4 \pi ^3}\nonumber\\
&&\times \Bigg\{ 2 E_W \sqrt{E_W^2-M_W^2} \left(M_W^2-6 E_W^2+8 E_W M_\chi-2 M_\chi^2\right)\nonumber\\
 &&+\Big(4 E_W^4-8 E_W^3 M_\chi+(2 E_W^2-M_W^2) \left(2M_\chi^2+ M_W^2\right)\Big) \nonumber\\
&&\times \ln\left[\frac{E_W+\sqrt{E_W^2-M_W^2}}{E_W-\sqrt{E_W^2-M_W^2}}\right]\Bigg\}.
\end{eqnarray}
The $W$ spectrum per
$\chi\chi\to e \nu W$ event is given in Fig.~\ref{dNdEW}. 
We use the scaling variable $x_W\equiv E_W/M_\chi$,
and plot 
$dN/dx_W \equiv (\frac{1}{\sigma_{e^+\nu W^-}}) \frac{d\sigma_{e^+\nu W^-}}{dx_W}$.
The kinematic range of $x_w$ is $[\frac{M_W}{M_\chi},\,(1+\frac{M_W^2}{4M_\chi^2}) ]$,
with the lower limit corresponding to a $W$ produced at rest, and the upper limit 
corresponding to parallel lepton momenta balancing the opposite W momentum.
As evident in Fig.~\ref{dNdEW}, the $W$ boson spectrum has a broad
energy distribution, including a significant component at high energy $E_W\sim M_\chi$.

\begin{figure}[t]
\includegraphics[width=1.0\columnwidth]{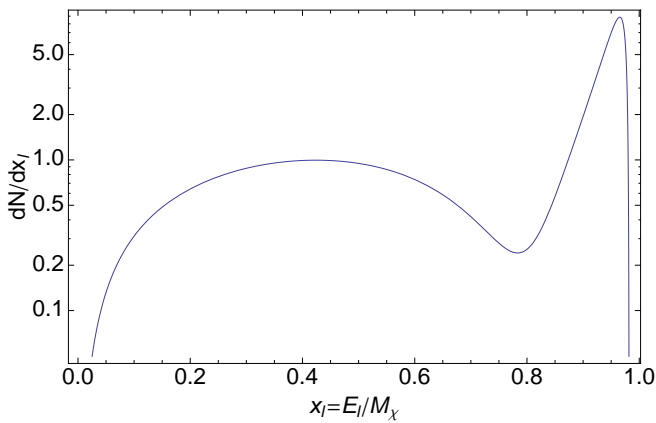}
\caption{
\label{dNdEe}
The primary lepton spectrum per $\chi\chi\to e \nu W$ for the example model, 
with $M_\chi=300$ GeV and $M_\eta^2\gg M_\chi^2$. 
}
\end{figure}

The energy spectrum of the the primary leptons is calculated in
similar fashion.  We present the analytic result in
Appendix~\ref{app:FullResults} (along with more detailed expressions
for $v\,\sigma$ and $v\,d\sigma/dE_W$).  Here the range of the scaling
variable $x_\ell\equiv E_\ell/M_\chi$ is
$[\,0,\,1-\frac{M_W^2}{4M_\chi^2} ]$.  Both limits arise when one lepton
has zero energy and the other is produced back-to-back with the $W$.
  The lepton spectrum is shown in Fig.~\ref{dNdEe}.  Note that this
  lepton spectrum is valid for either $e^+$ or $\nu$ from the
  annihilation $\chi\chi\to e^+ \nu W^-$, and for either $e^-$ or
  $\nubar$ from the annihilation $\chi\chi\to e^- \nubar W^+$.  
The primary lepton spectrum in
  Fig.~\ref{dNdEe} features a sharp cut off near $E_\ell=M_\chi$,
  and a dip in the spectrum that is due to an absorptive interference
  effect.

To obtain the full lepton spectrum, the contributions from the
subsequent decays of the gauge bosons to leptons must be included.
(The contribution from the lowest order $2\rarr 2$ process
$\chi\chi\to e^+e^- \ {\rm or\ }\nu\nubar$ is negligible.
We also neglect final state leptons resulting from $\mu$ decay and 
from the $\tau$ decay chain.
These leptons are softer than those we consider.)
For leptons from $W$-decay, the range of the scaling variable $x_\ell$ is
$[\frac{M_W^2}{4M_\chi^2},\, 1 ]$.
These limits arise when all four final state lepton momenta are collinear.
Particle spectra from the $W$~decay may be calculated in a simple but
approximate way, as we describe in Appendix~\ref{app:approxWdk} leading to Eq.~\rf{ellipse3}. 
The resulting secondary lepton spectrum is shown in
Fig~\rf{fig:SecLepSpectrum}.  Unsurprisingly, the spectrum of secondary leptons 
is softer than the spectrum of primary leptons.

When combining the primary lepton and secondary lepton spectra, the
relative weights are model dependent.  For example, the primary
$\ell$-spectrum is weighted by $BR(\chi\chi\rarr
W\nu\ell)+2BR(\chi\chi\rarr Z\ell\ell)$, while the secondary
$\ell$-spectrum is weighted by $BR(\chi\chi\rarr W+X)\times
BR(W\rarr\nu\ell) +BR(\chi\chi\rarr Z+X)\times BR(Z\rarr\ell\ell)$.

We note that the final charged-lepton spectra will by modified by
cosmic propagation effects.  The injected $e^\pm$ will suffer rapid
energy losses from synchrotron and inverse Compton processes on the
Universe's background magnetic and radiation fields (see, e.g.,
Ref.~\cite{Crocker:2010gy} for a recent analysis).  On the other hand,
the injected neutrinos do not interact with the environment, and so
their spectra remain unmodified.

\subsection{Unsuppressed Z Emission}

Consider the process producing the $\nubar\nu Z$ final state.  The
cross sections for the Z-strahlung processes are related to those for
W-strahlung in a simple way: The amplitudes producing $\nubar\nu Z$
arise from the same four graphs of Fig.~\rf{feyngraph1}, where $e$,
$W$ and $\eta^+$ are replaced everywhere by $\nu$ and $Z$ and
$\eta_0$, respectively.  The calculation of the amplitudes, and their
interferences, thus proceeds in an identical fashion.  After making
the replacement $M_W \rightarrow M_Z$, the cross section for the
annihilation process $\chi\chi\rightarrow\nu\bar \nu Z$ differs from
that for $\chi\chi\rightarrow e^+\nu W^-$ by only an overall
normalization factor,
\begin{eqnarray}
v\,\sigma_{\nu\bar\nu Z} &=& \left.\frac{1}{ (2\cos^2\theta_W )} \times
v\,\sigma_{e^+\nu W^-}\right|_{M_W\rightarrow M_Z} \nonumber\\ &\simeq&
0.65 \times v\,\sigma_{e^+\nu W^-} \Big|_{M_W\rightarrow M_Z} .
\end{eqnarray}

Consider now the $e^+e^-Z$ final state.  Again, the amplitudes arise
from the same four basic graphs of Fig.~\rf{feyngraph1}.  Since only
the left-handed leptons couple to the dark matter via the SU(2)
doublet $\eta$, only the left handed component of $e^-$ participates
in the interaction with the $Z$.  Therefore, the couplings of the
charged leptons to $Z$ and $W$
take the same form, up to a
normalization constant.  We thus find
\begin{eqnarray}
v\,\sigma_{e^+e^-Z}&=& 
\frac{2\left(\sin^2\theta_W - \frac{1}{2}\right)^2}{\cos^2\theta_W}
\times
v\,\sigma_{e^+\nu W^-} \, \Big|_{M_W\rightarrow M_Z}\nonumber\\
& \simeq & 0.19 \times \,v\,\sigma_{e^+\nu W^-}\Big|_{M_W\rightarrow M_Z}.
\end{eqnarray}

\section{Conclusions}
\label{sec:conclusion}

In an attempt to explain recent anomalies in cosmic ray data in a dark
matter framework, various non-standard properties have been invoked
such as dominant annihilation to leptons in so-called leptophilic
models.  When the dark matter is Majorana in nature, such
annihilations invariably are confronted by suppressions of such
processes via either p-wave velocity suppression or helicity
suppression.  With the aid of Fierz transformation technology, which
we have presented in some detail, we have elucidated the general
circumstances where suppressions may be encountered.

It has been known for some time that photon bremsstrahlung may have a
dramatic effect on such suppressions.  We have shown that once one
considers the inclusion of three body final states due to electroweak
bremsstrahlung, one may also lift these suppressions and obtain rates
which may be several orders of magnitude beyond those without such
radiative corrections.  In fact, barring an unexpected mass-degeneracy, 
the EW-bremsstrahlung lifts the suppression at one order lower in a
certain small ratio of squared masses than does EM-bremsstrahlung,
as explained in the text.

Such radiative processes may be lethal for models attempting to
produce positrons without overproducing antiprotons due to the
subsequent hadronization of the radiated gauge bosons.
Given that electroweak bremsstrahlung is the dominant annihilation
channel for the DM models under consideration, and both $W$ and $Z$
decay to hadrons with a branching ratio of approximately 70\%, a large
hadronic component is unavoidable.  Importantly in the context of
recent cosmic ray data, there will be sizable antiproton production.
We also note that dark matter searches triggering on anti-deuterons
will find a sample in the $W$- and $Z$-bremsstrahlung processes.  The
Aleph experiment has measured an anti-deuteron production rate of
$5.9\pm1.9\times 10^{-6}$~anti-deuterons per hadronic decay of the
Z~\cite{anti-deuteron}.  We expect the rate for anti-deuteron
production in $W$-decay to be similar.

Even for models which do not suffer a suppression of the lowest order
process, we see that it is impossible to have purely leptonic
annihilation products, including ``leptophilic'' models in which the
dark matter has direct couplings only to leptons.  In a broader
context the results presented here show the importance that may be
played by electroweak bremsstrahlung in future searches of indirect
dark matter detection.  For any DM model for which electroweak
bremsstrahlung makes an important contribution to observable fluxes,
there will be large, correlated fluxes of $e^\pm$, neutrinos, hadrons
and gamma rays.  We will explore the detection of these signals in a
future article~\cite{BDJWnext}.

\section*{Acknowledgements}
We thank Sheldon Campbell, Bhaskar Dutta, Sourish Dutta, Haim Goldberg, Lawrence Krauss, 
Danny Marfatia, Yudi Santoso and Nick Setzer for helpful discussions.  
NFB was supported by the
Australian Research Council, TDJ was supported by the Commonwealth of
Australia, and TJW and JBD were supported in part by U.S.~DoE grant
DE--FG05--85ER40226.  TJW benefited from a AvHumboldt Senior Research
Award and hospitality at MPIK (Heidelberg), MPIH (Munich), U\@. Dortmund and
the Aspen Center for Physics.

\appendix

\section{Fundamentals of Fierzing}
\label{app:Fierz}

\noindent
In this paper we have made use of standard Fierz transformations, helicity-basis Fierz 
transformations,
and generalizations of the two. In this Appendix, we derive these transformations. 
The procedure for standard Fierz transformation can be found in,
e.g.,~\cite{IZp161-2}, while more general Fierz transformations are
laid down in~\cite{Nishi:2004st}.  The starting point is to define a
basis $\{\Gamma^B\}$ and a dual basis $\{\Gamma_B\}$, each spanning
$4\times 4$ matrices over the complex number field ${\cal C}$, such
that an orthogonality relation holds.  The standard Fierz
transformation uses the ``hermitian'' bases
\bea{stdbasis}
\{\Gamma^B\}&=&\{\openone, i\gamma_5,\gamma^\mu,\gamma_5\gamma^\mu,\sigma^{\mu\nu}\}\,,
\quad {\rm and}\quad \nonumber \\
\{\Gamma_B\}&=&\{\openone, (-i\gamma_5),\gamma_\mu,(-\gamma_5\gamma_\mu),\half\sigma_{\mu
\nu}\}\,,
\eea
respectively.
Because of their Lorentz and parity transformation properties, these basis matrices and their duals  
are often labeled as $S$ and $\tS$ (scalars), $P$ and $\tP$ (pseudoscalars), 
$V$ and $\tV$ (vectors, four for $V$, four for $\tV$), $A$ and $\tA$ (axial vector, four for $A$, four for 
$\tA$), 
and $T$ and $\tT$ (antisymmetric tensor, six for $T$, six for $\tT$).
As usual, spacetime indices are lowered with the Minkowski metric,
$\gamma_5=\gamma^5=i\gamma^0\gamma^1\gamma^2\gamma^3$, 
$\sigma^{\mu\nu}\equiv \frac{i}{2}\,[\gamma^\mu,\gamma^\nu ]$,
(and $\gamma^5\sigma^{\mu\nu}=\frac{i}{2}\epsilon^{\mu\nu\alpha\beta}\sigma_{\alpha\beta}$).
Note the change of sign between the the basis and dual for the P and A matrices.
The bases are ``hermitian'' in that $\gamma^0\,\Gamma_B^\dag \gamma^0=\Gamma_B$,
so that the associated Dirac bilinears satisfy 
$[{\bar\Psi}_1 \Gamma^B\Psi_2]^\dag = {\bar\Psi}_2 \Gamma^B\Psi_1$ and 
$[{\bar\Psi}_1 \Gamma_B\Psi_2]^\dag = {\bar\Psi}_2 \Gamma_B\Psi_1$. 
Importantly, we have $\Gamma_B=(\Gamma^B)^{-1}$ in the sense of the accompanying 
orthogonality relation 
\beq{ortho1}
Tr\,[\Gamma_C\,\Gamma^B\,]=4\,\delta^B_C\,, 
\quad B,\,C = 1,\dots, 16\,.
\eeq
Note that the factor of $\half$ in the definition of $\tT=\half\sigma_{\mu\nu}$ 
(but not in $T=\sigma^{\mu\nu}$)
provides the normalization required by Eq.~\rf{ortho1}:
\beq{normzn1}
Tr\,[\Gamma_B\,\Gamma^B]_{({\rm no sum})} = \sum_C\,Tr\,[\Gamma_C\,\Gamma^B] = 4\,.
\eeq

The orthogonality relation allows us to expand any $4\times 4$ complex matrix $\X$ in terms of the 
basis as 
\bea{expn1}
\X &=& \X_B\,\Gamma^B=\X^B\,\Gamma_B\,,\quad {\rm with} \quad \nonumber\\
\X_B&=&\frac{1}{4} Tr\,[\,\X\,\Gamma_B\,]\,, \ {\rm and}\ \X^B=\frac{1}{4} Tr\,[\,\X\,\Gamma^B\,]\,, 
\nonumber\\
{\rm i.e.,}\quad \X &=& \quarter\,Tr\,[\X\,\Gamma^B]\,\Gamma_B = \quarter\,Tr\,[\X\,\Gamma_B]\,
\Gamma^B\,.
\eea
One readily finds that the particular matrix element $(\X)_{ab}$ satisfies
\beq{arb}
(\X)_{cd}\,\delta_{db}\,\delta_{ac}=\quarter\,[(\X)_{cd}\,(\Gamma_B)_{dc}\,]\,(\Gamma^B)_{ab}\,.
\eeq
Since each element $(\X)_{cd}$ is arbitrary, Eq.~\rf{arb} is equivalent to a completeness relation
\beq{completeness1}
(\openone)\,[\openone] = \quarter\,(\Gamma_B\,]\,[\Gamma^B\,) = \quarter\,(\Gamma^B\,]\,
[\Gamma_B\,)\,,
\eeq
where we have adopted Takahashi's notation~\cite{Taka1986} where matrix indices are replaced 
by parentheses
$(\cdots)$ and brackets $[\cdots]$, in an obvious way.
Thus, any $4\times 4$ matrices $\X$ and $\Y$ may be expressed as 
\bea{XandY1}
(\X)\,[\Y] &=& (\X\openone)\ [\openone \Y]=
    \frac{1}{4}\,(\X\,\Gamma^B\,\Y\,]\ [\,\Gamma_B\,) \nonumber\\ 
&=& \frac{1}{4^2}\,Tr\,[\X\,\Gamma^B\,\Y\,\Gamma_C ]\ (\Gamma^C\,]\ [\Gamma_B\,)\,.
\eea
This equation is presented as Eq.~\rf{genFierz} in the main text.
Alternatively, we may express any $4\times 4$ matrices $\X$ and $\Y$ as 
\bea{XandY2}
(\X)\,[\Y] &=& (\X\openone)\ [Y\openone]= 
    \frac{1}{4}\,(\X\,\Gamma^B]\ [Y\,\Gamma_B\,) \\
 &=&  \frac{1}{4^3}\,Tr\,[\X\,\Gamma^B\,\Gamma_C ]\ Tr\,[\Y\,\Gamma_B\,\Gamma^D ]\ (\Gamma^C\,]\ 
[\Gamma_D\,)\,.\nonumber
\eea

The RHS's of Eqs.~\rf{XandY1} and \rf{XandY2} offer two useful options for Fierzing matrices.
The first option sandwiches both LHS matrices into one of the two spinor bilinears,
and ultimately into a single long trace.
The second option sandwiches each LHS matrix into a separate spinor bilinear,
and ultimately into separate trace factors.
Eq.~\rf{XandY1} seems to be more useful than \rf{XandY2}.
One use we will make of Eq.~\rf{XandY1} will be to express chiral vertices in terms of Fierzed 
standard vertices.
But first we reproduce the standard Fierz transformation rules by setting $\X=\Gamma^D$ and $\Y=
\Gamma_E$ in Eq.~\rf{XandY1},
to wit:
\beq{stdFierz1}
(\Gamma^D )\,[\Gamma_E ]=\frac{1}{4^2}\,Tr\,[\Gamma^D\,\Gamma^B\,\Gamma_E\,\Gamma_C ]\,
(\Gamma^C ]\,[\Gamma_B )\,.
\eeq
(An additional minus sign arises if the matrices are sandwiched between anticommuting field 
operators, 
rather than between Dirac spinors.)
Evaluation of the trace in Eq.~\rf{stdFierz1} for the various choices of $(B,C)$ leads to the oft-quoted
result~\cite{IZp161-2}
\beq{stdFierz2}
\left(
\barr{c}
(S)\ [\tS] \\
(V)\ [\tV] \\
(T)\ [\tT] \\
(A)\ [\tA] \\
(P)\ [\tP] 
\earr
\right)
= \quarter
\left(
\barr{rrrrr}
 1 & 1 & 1 & 1 & 1 \\
 4 &-2 & 0 & 2 &-4 \\
 6 & 0 &-2 & 0 & 6 \\
 4 & 2 & 0 &-2 &-4 \\
 1 &-1 & 1 &-1 & 1 
\earr
\right)
\ \left(
\barr{c}
(S]\ [\tS) \\
(V]\ [\tV) \\
(T]\ [\tT) \\
(A]\ [\tA) \\
(P]\ [\tP) 
\earr
\right)\,.
\eeq
More relevant for us, as will be seen, is the ordering $P,\,S,\,A,\,V,\,T$, which leads to a
Fierz matrix obtained from the one above with the swapping of matrix indices 
$1\rarr 2\rarr 4\rarr 3\rarr 5\rarr 1$.
The result is
\beq{stdFierz3}
\left(
\barr{c}
(P)\ [\tP] \\
(S)\ [\tS] \\
(A)\ [\tA] \\
(V)\ [\tV] \\
(T)\ [\tT] 
\earr
\right)
= \quarter
\left(
\barr{rrrrr}
 1 & 1 &-1 &-1 & 1 \\ 
 1 & 1 & 1 & 1 & 1 \\
-4 & 4 &-2 & 2 & 0 \\
-4 & 4 & 2 &-2 & 0 \\
 6 & 6 & 0 & 0 &-2 
\earr
\right)
\ \left(
\barr{c}
(P]\ [\tP) \\
(S]\ [\tS) \\
(A]\ [\tA) \\
(V]\ [\tV) \\
(T]\ [\tT) 
\earr
\right)\,.
\eeq
(The zeroes make it clear that Fierzing induces no coupling between tensor interactions and vector
and axial vector interactions.) As an example of how to read this matrix, 
\begin{equation}
(A)\ [\tA] = -(P]\ [\tP)+(S]\ [\tS)- \frac{1}{2}(A]\ [\tA)+\frac{1}{2}(V]\ [\tV),
\end{equation}
or, multiplying by spinors and giving the explicit forms of the gamma-matrices,
\begin{eqnarray}
&& \left(\ubar  \gamma_5 \gamma^\mu u\right) \ \left(\vbar (-\gamma_5 \gamma_\mu) v \right) 
\nonumber \\
&& \;\;\; =  -\left(\ubar i\gamma_5 v\right)\ \left(\vbar \left(-i\gamma_5 \right)u\right)+\left(\ubar  v\right)\ \left(\vbar u \right) \\
&& \;\;\; - \frac{1}{2}\left(\ubar \gamma_5 \gamma^\mu v\right)\ \left(\vbar \left(-\gamma_5\gamma_\mu
\right) u\right)+\frac{1}{2}\left(\ubar \gamma^\mu v\right)\ \left(\vbar \gamma_\mu u\right).\nonumber  
\end{eqnarray}

The Fierz matrix $M$ for the standard basis is nonsingular, and hence has five nonzero 
eigenvalues $\lambda_j$.
Since two swaps of Dirac indices returns the indices to their original order,
the matrix is idempotent, with $M^2=\openone$, or equivalently, $M^{-1}=M$.
Accordingly, the five eigenvalues satisfy $\lambda_j^2=1$, so individual eigenvalues 
must be $\lambda_j=\pm 1$.
Also, the corresponding eigenvectors are invariant under the interchange of two Dirac indices.
In Table~\rf{FierzInvariants1} we list the eigenvalues and ``Fierz-invariant'' eigenvectors.

\begin{table}[tdp]
\caption{Fierz-invariant combinations in the standard basis.}
\begin{center}
\begin{tabular}{|c|c|}
\hline
Fierz-invariant combination & eigenvalue \\
\hline\hline
$ 3\ (S\otimes\tS + P\otimes\tP) +  T\otimes\tT$ & $+1$ \\ \hline
$ 2\ (S\otimes\tS - P\otimes\tP) + (V\otimes\tV +A\otimes\tA)$ & $+1$ \\
\hline\hline
$ V\otimes\tV - A\otimes\tA$ & $-1$ \\ \hline
$ S\otimes\tS + P\otimes\tP - T\otimes\tT$ & $-1$ \\ \hline
$ 2\ (S\otimes\tS - P\otimes\tP) - (V\otimes\tV + A\otimes\tA)$ & $-1$ \\ \hline
\end{tabular}
\end{center}
\label{FierzInvariants1}
\end{table}

Helicity projection operators are often present in theories where the DM couple to the $SU(2)$ 
lepton doublet,
so it is worth considering Fierz transformations in the more convenient chiral basis.

One derivation of chiral Fierz transformations utilizes the following 
chiral bases~(hatted)~\cite{Nishi:2004st}:
\bea{chiralbasis2}
\{\Ghat^B\}&=&\{P_R,\,P_L,\,P_R\gamma^\mu,\,P_L\gamma^\mu,\,\half\sigma^{\mu\nu} \}\,,
\quad {\rm and} \quad \nonumber\\
\{\Ghat_B\}&=&\{P_R,\,P_L,\,P_L\gamma_\mu,\,P_R\gamma_\mu,\,\half\,\sigma_{\mu\nu} \}\,,
\eea
where $P_R\equiv\half (1+\gamma_5)$ and $P_L\equiv\half (1-\gamma_5)$ 
are the usual helicity projectors.
The orthogonality property between the chiral basis and its dual is
\beq{ortho2}
Tr\,[\Ghat_C\,\Ghat^B\,]=2\,\delta^B_C\,,\quad B,\,C = 1,\dots, 16\,,
\eeq
which implies the normalization
\beq{normzn2}
Tr\,[\Ghat_B\,\Ghat^B]_{({\rm no\ sum})} = \sum_C\,Tr\,[\Ghat_C\,\Ghat^B] = 2\,.
\eeq
Notice that because $\{\gamma_5,\gamma^\mu\}=0$, the dual of $P_R\gamma^\mu$ 
is $P_L\gamma_\mu$, and the dual of $P_L\gamma^\mu$ is $P_R\gamma_\mu$.
Notice also that the normalization for the chiral bases necessitates factors of 
$\half$ in both $\hT=\half\sigma^{\mu\nu}$ and $\tilde{\hT}=\half\sigma_{\mu\nu}$, 
in contrast to the tensor elements of the standard bases, given in Eq.~\rf{stdbasis}.

In the chiral basis, one is led to a general expansion 
\beq{chiralX}
 \X = \half\,Tr\,[\X\,\Ghat^B]\,\Ghat_B = \half\,Tr\,[\X\,\Ghat_B]\,\Ghat^B\,,
\eeq
and to a completeness relation
\beq{completeness2}
(\openone)\,[\openone] = \half\,(\Ghat_B\,]\,[\Ghat^B\,) = \half\,(\Ghat^B\,]\,[\Ghat_B\,)\,.
\eeq
Thus, any $4\times 4$ matrices $\X$ and $\Y$ may be expressed as 
\bea{chiralXandY}
(\X)\,[\Y] &=& (\X\openone)\,[\openone \Y]=\half (\X\,\Ghat_B\,\Y\,]\,[\,\Ghat^B\,) \nonumber\\
   &=& \quarter\,Tr\,[\X\,\Ghat^C\,\Y\,\Ghat_B ]\ (\Ghat^B\,]\,[\Ghat_C\,)\,.
\eea
Substituting $\X=\Ghat^D$ and $\Y=\Ghat_E$ into Eq.~\rf{chiralXandY}, one gets
\beq{}
(\Ghat^D )\,[\Ghat_E ]=\frac{1}{4}\,Tr\,[\Ghat^D\,\Ghat^C\,\Ghat_E\,\Ghat_B ]\ (\Ghat^B ]\;[\Ghat_C )\,.
\eeq
Evaluating the trace in Eq.~\rf{chiralFierz1} leads to the chiral-basis 
analog of \rf{stdFierz2} or \rf{stdFierz3}, presented in Eq.~\rf{chiralFierz} of the main text.

As a check, we note that the matrix $M$ in Eq.~\rf{chiralFierz} is idempotent, $M^2= \openone$,
as it must be.
The eigenvalues are therefore $\pm1$.
Eigenvalues and Fierz-invariant eigenvectors for the chiral basis are given in Table~\rf
{FierzInvariants2}.
The final two eigenvectors in the Table simply express again the invariance of $V\pm A$ 
interactions under 
Fierz-transposition of Dirac indices.
This invariance is also evident in the diagonal nature of the bottom two rows of 
the matrix Eq.~\rf{chiralFierz}.

\begin{table}[tbp]
\caption{Fierz-invariant combinations in the chiral basis.}
\begin{center}
\begin{tabular}{|c|c|}
\hline
Fierz-invariant combination & eigenvalue \\
\hline\hline
$ 3\ (P_R\otimes P_R + P_L\otimes P_L)+\hT\otimes{\tilde\hT}$ & $+1$ \\ \hline
$ 2\ P_R\otimes P_L + P_R\gamma^\mu\otimes P_L\gamma_\mu$     & $+1$ \\ \hline
$ 2\ P_L\otimes P_R + P_L\gamma^\mu\otimes P_R\gamma_\mu$     & $+1$ \\ \hline
\hline
$ P_R\otimes P_R + P_L\otimes P_L - \hT\otimes{\tilde\hT}$    & $-1$ \\ \hline
$ 2\ P_R\otimes P_L - P_R\gamma^\mu\otimes P_L\gamma_\mu$     & $-1$ \\ \hline
$ 2\ P_L\otimes P_R - P_L\gamma^\mu\otimes P_R\gamma_\mu$     & $-1$ \\ \hline
$ P_R\gamma^\mu\otimes P_R\gamma_\mu$                     & $-1$ \\ \hline
$ P_L\gamma^\mu\otimes P_L\gamma_\mu$                     & $-1$ \\ \hline

\end{tabular}
\end{center}
\label{FierzInvariants2}
\end{table}

One may instead want the Fierz transformation that takes chiral bilinears to standard bilinears.
Since models are typically formulated in terms of chiral fermions, a projection onto standard 
$s$-channel bilinears would be well- suited for a partial wave analysis.
Because different partial waves do not interfere with one another, the calculation simplifies in terms of 
$s$-channel partial waves.

Setting $\X=\Ghat^D$ and $\Y=\Ghat_E$ in Eq.~\rf{XandY1}, we readily get
\beq{nonstdFierz1}
(\Ghat^D )\,[ \Ghat_E ]=\frac{1}{4^2}\,Tr\,[ \Ghat^D\,\Gamma^B\,\Ghat_E\,\Gamma_C ]\,(\Gamma^C ]\,
[ \Gamma_B )\,.
\eeq
We (should) get the same result by resolving the RHS vector in Eq.~\rf{chiralFierz} into standard 
basis matrices.
The result is 
\begin{widetext}
\bea{chiralTOstd}
\left(
\barr{c}
(P_R)\ [P_R] \\ 
(P_L)\ [P_L] \\
(P_R\gamma^\mu)\ [P_L\gamma_\mu] \\  
(P_L\gamma^\mu)\ [P_R\gamma_\mu] \\
(\hT)\ [\hT] \\
\!\! (\gamma_5\hT)\ [\hT] \\
(P_R)\ [P_L] \\ 
 (P_L)\ [P_R] \\
(P_R\gamma^\mu)\ [P_R\gamma_\mu] \\
(P_L\gamma^\mu)\ [P_L\gamma_\mu] \\
\earr
\right)
= \frac{1}{8}
\left(
\barr{rrrrrr|rrrr}
 1 & 1 & 1 & 1 & 1 & 1 & 0 & 0 & 0 & 0   \\ 
 1 & 1 &-1 &-1 & 1 &-1 & 0 & 0 & 0 & 0   \\
 4 &-4 & 4 &-4 & 0 & 0 & 0 & 0 & 0 & 0   \\
 4 &-4 &-4 & 4 & 0 & 0 & 0 & 0 & 0 & 0   \\ 
 6 & 6 & 0 & 0 &-2 & 0 & 0 & 0 & 0 & 0   \\
 0 & 0 & 6 & 6 & 0 & 2 & 0 & 0 & 0 & 0 \\  \hline
 0 & 0 & 0 & 0 & 0 & 0 & 1 &-1 & 1 &-1   \\ 
 0 & 0 & 0 & 0 & 0 & 0 & 1 &-1 &-1 & 1   \\ 
 0 & 0 & 0 & 0 & 0 & 0 &-2 &-2 &-2 &-2   \\
 0 & 0 & 0 & 0 & 0 & 0 &-2 &-2 & 2 & 2   \\
\earr
\right)
\left(
\barr{c}
(\openone]\      [\openone)  \\
(\gamma_5]\      [\gamma_5) \\
(\gamma_5]\      [\openone) \\
(\openone]\      [\gamma_5) \\
       (T]\      [\tilde{T})      \\
\!\! (\gamma_5 T]\  [\tilde{T})   \\
(\gamma^\mu]\    [\gamma_\mu) \\
(\gamma_5\gamma^\mu]\ [\gamma_5\gamma_\mu) \\
\!\! (\gamma_5\gamma^\mu]\ [\gamma_\mu) \\
\ (\gamma^\mu]\    [\gamma_5\gamma_\mu) \\
\earr
\right)
\eea
\end{widetext}
All relations are invariant under the simultaneous interchanges 
$P_R\leftrightarrow P_L$ and $\gamma_5\rarr -\gamma_5$.
The matrix in \rf{chiralTOstd}, relating two different bases, is not idempotent.  
In fact, it is singular.

\section{Cancellation of Vector and Tensor Amplitudes for Majorana Fermions}
\label{app:VectorTensor}

Majorana particles are invariants under charge conjugation ${\cal C}$.
Accordingly, the Majorana field creates and annihilates the same particle.
This implies that for each $t$-channel diagram, there is an accompanying $u$-channel diagram,
obtained by interchanging the momentum and spin of the two Majorana fermions.
The relative sign between the $t$- and $u$-channel amplitudes is $-1$ in accord with Fermi statistics.
For example, consider the Fierzed (i.e., $s$-channel) bilinear for $\chi$-annihilation:
$\bar{v}(k_1,s_1) \Gamma_B u(k_2,s_2)$.
The associated Fierzed bilinear from the $(k_1\leftrightarrow k_2)$-exchange graph, 
with its relative minus sign, is 
$-\bar{v}(k_2,s_2) \Gamma_B u(k_1,s_1)$.
Constraints relating the four-component Dirac
spinors to their underlying two-component Majorana spinors must be imposed.  
These constraints, any one of which implies the other three, are
\bea{Majspinors}
&& u(p,s) = C\bar{v}^T(p,s)\,,\quad  \bar{u}(p,s)=-v^T(p,s)C^{-1}\,, \nonumber \\
&& v(p,s) = C\bar{u}^T(p,s)\,,\quad 
\bar{v}(p,s) = -u^T(p,s)C^{-1}\,.
\eea
Here, $C$ is the charge conjugation matrix. 
These Majorana conditions on the spinors allow us to rewrite the exchange bilinear as 
(suppressing spin labels for brevity of notation)
\begin{eqnarray}
\label{Majswap}
-\bar{v}(k_2)\Gamma_B u(k_1)) &=& u^T(k_2)C^{-1}\Gamma_B C\bar{v}^T(k_1)\nonumber \\
&=&\left[ \bar{v}(k_1)(C^{-1}\Gamma_B C)^T u(k_2)\right]^T \nonumber \\
&=& \bar{v}(k_1) (\eta_B\Gamma_B) u(k_2)\,.
\end{eqnarray}
For the final equality, we have used (i) the fact that the transpose symbol can be dropped from a number,
and (ii) the identity $(C^{-1}\Gamma_B C)^T=(\eta_B (\Gamma_B)^T)^T =\eta_B\Gamma_B$,
where  $\eta_B=+1$ for $\Gamma =$ scalar, pseudoscalar, axial vector, and 
$\eta_B = -1$ for $\Gamma =$ vector or tensor.

In the Four-Fermi or heavy propagator limit, where the differing
momenta in the $t$- and $u$-channel propagators can be ignored, one
obtains an elegant simplification.  
Subtracting the u-channel amplitude from the t-channel amplitude, one
arrives at the weighting factor $(1+\eta_B)$, which is two for $S,P,$
and $A$ couplings, and zero for $V$ and $T$ couplings.  Thus, we must
drop $V$ and $T$ couplings appearing in the Fierzed bilinears of the
$\chi$-current.  What this means for the model under discussion is
that after Fierzing, only the axial vector coupling of the
$\chi$-current remains, and the factor of $1+\eta_A=2$ is multiplied
by the (7-8)-element~$=\half$ in the Fierz matrix of
Eq.~\rf{chiralFierz} to give a net weight of~1.

\section{Non-Relativistic and Extreme-Relativistic Limits of Fermion Bilinears}
\label{app:NRnERlimits}
We work in the chiral representation of the Dirac algebra,
and we follow the notation of~\cite{Peskin}.
Accordingly,
\beq{chiralDirac}
\gamma_0=
\left(
\barr{rr}
  0         & \unitmatrix \\ 
\unitmatrix & 0 
\earr
\right)\,,
\quad
{\vec\gamma}=
\left(
\barr{rr}
  0           & {\vec\sigma} \\
-{\vec\sigma} & 0 
\earr
\right)\,,
\quad
\gamma_5 =
\left(
\barr{rr}
-\unitmatrix & 0 \\ 
  0          & \unitmatrix 
\earr
\right)\,.
\eeq
The rest-frame four-spinor is 
\beq{RFspinor}
u({\vec p}=0)=\sqrt{M}
\left(
\barr{c}
\xi \\
\xi
\earr
\right)\,,
\eeq
where
$\xi$ is a two-dimensional spinor.
The spinor with arbitrary momentum is obtained by boosting.
One gets
\beq{arbuspinor}
u(p)=
\left(
\barr{c}
\sqrt{p\cdot\sigma}\;\xi \\
\\
\sqrt{p\cdot\bsigma}\;\xi
\earr
\right)\,,
\eeq
where $\sigma\equiv (1,\vec{\sigma})$ and $\bsigma\equiv (1,-\vec{\sigma})$.

In a standard fashion, we choose the up and down spin eigenstates of $\sigma_3$ as the basis
for the two-spinors. These basis two-spinors are 
\beq{spinorbasis}
\xi_+\equiv
\left(
\barr{c}
 1 \\
 0 
\earr
\right)\,,
\quad
\xi_-\equiv
\left(
\barr{c}
 0 \\
 1 
\earr
\right)\,.
\eeq
In terms of the chosen basis, we have for the NR $u$-spinors,
\beq{NRu}
u_\pm\stackrel{NR}{\longrarr} \sqrt{M}
\left(
\barr{c}
\xi_\pm \\
\xi_\pm
\earr
\right)\,.
\eeq
We get the ER limit of the $u$-spinors from Eq.~\rf{arbuspinor}.
After a bit of algebra, one finds 
\beq{ERu}
u_+\stackrel{ER}{\longrarr} \sqrt{2E}
\left(
\barr{c}
0 \\
0 \\
\xi_+
\earr
\right)\,,
\quad
u_-\stackrel{ER}{\longrarr} \sqrt{2E}
\left(
\barr{c}
\xi_- \\
0 \\
0
\earr
\right)\,.
\eeq

The arbitrary $v$-spinor is given by
\beq{arbvspinor}
v(p)=
\left(
\barr{r}
  \sqrt{p\cdot\sigma}\;\eta \\
\\
- \sqrt{p\cdot\bsigma}\;\eta
\earr
\right)\,.
\eeq
In the Dirac bilinear the two-spinor $\eta$ is independent of the two-spinor $xi$,
and so it is given an independent name, $\eta$.  However, the basis $\eta^\pm$ remains $\xi^\pm$ as defined above.
It is the minus sign in the lower components of $v$ relative to the upper components 
that distinguishes $v$ in eq.~(\ref{arbvspinor}) from $u$ in eq.~(\ref{arbuspinor}) in a fundamental way.
After a small amount of algebra, one finds the limits
\beq{NRv}
v_\pm\stackrel{NR}{\longrarr} v_\pm({\vec p}=0) = \sqrt{M}
\left(
\barr{r}
  \eta_\pm \\
- \eta_\pm
\earr
\right)\,,
\eeq
and
\beq{ERv}
v_+\stackrel{ER}{\longrarr} \sqrt{2E}
\left(
\barr{c}
  0 \\
  0 \\
-\eta_+
\earr
\right)\,,
\quad
v_-\stackrel{ER}{\longrarr} \sqrt{2E}
\left(
\barr{c}
\eta_- \\
 0 \\
 0
\earr
\right)\,.
\eeq

Finally, we apply the above to determine the values of Dirac bilinears in the NR and ER limits.
The $\bar{u}\equiv u^\dag\,\gamma_0$ and $\bar{v}\equiv v^\dag\,\gamma_0$ conjugate spinors are are easily found 
from the $u$ and $v$ spinors.
We let $\Gamma$ denote any of the hermitian basis Dirac-matrices 
$\{\unitmatrix,\ i\,\gamma_5,\ \gamma^\mu,\ \gamma_5\,\gamma^\mu,\ \sigma^{\mu\nu}\}$.
Then, the NR limit of $\bar{u}(p_1)\,\Gamma\,v(p_2)$ is just 
\beq{NRuGp}
\bar{u}(p_1)\,\Gamma\,v(p_2)\stackrel{NR}{\longrarr}M\,
\left[
\, (\xi_1,\ \xi_1)\ \Gamma\,
\left(
\barr{r}
 \eta_2 \\
-\eta_2
\earr
\right)
\right] \,.
\eeq
Non-relativistic results for the various choices of basis $\Gamma$'s and spin combinations 
are listed in Table~\ref{table:bilinearlimits} of the text.

To give a succinct formula for the ER limit of $\bar{u}(p_1)\,\Gamma\,v(p_2)$,
we take ${\hat p_1}=-{\hat p_2}= {\hat 3}$, i.e. we work in a frame where ${\hat p_1}$ and ${\hat p_2}$ are collinear,
and we quantize the spin along this collinear axis.
The result is 
\beq{ERuGp}
\bar{u}(p_1)\,\Gamma\,v(p_2)\stackrel{ER}{\longrarr}\sqrt{4E_1 E_2}\ \ 
\left[
\, \xi_1\ \ (\Lambda_+,\,\Lambda_- )\,
    \Gamma\,
\left(
\barr{r}
 \Lambda_+ \\
-\Lambda_-
\earr
\right)
\ \eta_2\,\right] \,,
\eeq
where the matrices $\Lambda_\pm$ are just up and down spin projectors along the quantization axis ${\hat 3}$:
\beq{Lambdas}
\lambda_+ = 
\left(
\barr{cc}
 1 & 0 \\
 0 & 0 
\earr
\right)\,,
\quad
\Lambda_- = 
\left(
\barr{cc}
 0 & 0 \\
 0 & 1 
\earr
\right)\,.
\eeq
Extreme-relativistic results for the various choices of basis $\Gamma$'s and spin combinations are listed in Table~\ref{table:bilinearlimits}
of the text.

\section{Full Cross Section Results}
\label{app:FullResults}

We present here the full results of the cross section calculations for
the process $\chi\chi\rightarrow\e^\mp\overset{(-)}{\nu} W^\pm$,
including terms of all orders in $M_\chi$.  In Section
~\ref{sec:example} we presented only the leading order terms, 
which dominate in the large $M_\chi$ limit.  
For $M_\chi$ not too much heavier than $M_W$, 
it is important to retain sub-leading terms.

The total cross section for $\chi\chi\rightarrow\e^\mp\overset{(-)}{\nu} W^\pm$ is given by  

\begin{widetext}
\begin{align}
v\,\sigma_{\e^+\nu W^-} = &\frac{g^2 f^4}{2^{13} M_W^2 M_\eta^4 \pi ^3}
\Bigg\{  \left(\frac{7}{32} \frac{M_W^8}{M_\chi^4} - \frac{7}{9}  \frac{M_W^6}{M_\chi^2}+4 M_W^4-16 M_W^2 M_\chi^2-\frac{56}{9} M_\chi^4\right)  
\nonumber\\
&+ \ln\left[\frac{4 M_\chi^2}{M_W^2}\right] 
\bigg(\frac{1}{16} \frac{M_W^8}{M_\chi^4}+ \frac{4}{3} \frac{M_W^6}{M_\chi^2} - 2 M_W^4 + 16 M_W^2 M_\chi^2 +\frac{16}{3} M_\chi^4 
+ 8 M_W^2 \left(M_W^2+2 M_\chi^2\right) \ln\left[\frac{2 M_W M_\chi}{M_W^2+4 M_\chi^2}\right]\bigg)
\nonumber\\
&+ 8 M_W^2 \left(M_W^2+2 M_\chi^2\right) \left(\text{Li}_2\left[\frac{4 M_\chi^2}{M_W^2+4 M_\chi^2}\right]-\text{Li}_2\left[\frac{M_W^2}{M_W^2+4 M_\chi^2}\right]\right)
+\Order(v^2,M_\eta^{-2},m_\ell^2) \Bigg\}.\label{vsigmaFull}
\end{align}
The $W$ energy spectrum is
\begin{align}
\frac{v\,d\sigma_{\e^+\nu W^-}}{dE_W}
&=\frac{g^2 f^4}{512 E_W M_W^2 M_\eta^4 \pi ^3}
\Bigg\{ 2 E_W \sqrt{E_W^2-M_W^2} \left(M_W^2-6 E_W^2+8 E_W M_\chi-2 M_\chi^2\right)
\\
&+\Big(4 E_W^4-8 E_W^3 M_\chi+(2 E_W^2-M_W^2) \left(2M_\chi^2+ M_W^2\right)\Big) 
\ln\left[\frac{E_W+\sqrt{E_W^2-M_W^2}}{E_W-\sqrt{E_W^2-M_W^2}}\right]
+\Order(v^2,M_\eta^{-2},m_\ell^2)\Bigg\},
\nonumber
\end{align}
while the lepton spectrum (for either the charged lepton or the neutrino)  is
\begin{align}
\frac{v\,d\sigma_{\e^+\nu W^-}}{dE_\ell}
&=\frac{g^2f^4}{2^{18} M_\eta^4 (M_\chi-E_e) \pi ^3}\Bigg\{\frac{E_e \left(4 (M_\chi-E_e)
M_\chi-M_W^2\right)}{M_W^2 M_\chi^4 \left(M_W^2+4 E_e M_\chi\right) (M_\chi-E_e)^4}
\nonumber\\&
\times\bigg(7\times 2^8 E_e^7 M_W^4
+2^8 E_e^6 M_\chi^3 \left(M_W^2-\frac{139}{3}M_\chi^2\right)
-2^5 E_e^5 M_\chi^2 \left(M_W^4+\frac{146}{3}M_\chi^2 M_W^2-984 M_\chi^4\right)
\nonumber\\&
-2^4 E_e^4 M_\chi \left(M_W^6-13 M_W^4 M_\chi^2-232 M_\chi^4 M_W^2+2704 M_\chi^6\right)
\nonumber\\&
+E_e^3 \left(-M_W^8+\frac{164}{3}M_W^6 M_\chi^2-560 M_W^4 M_\chi^4-4672 M_\chi^6
M_W^2+\frac{191\times 2^9}{3}M_\chi^8\right)
\nonumber\\&
+8 E_e^2 M_\chi \left(\frac{1}{3}M_W^8-6 M_W^6 M_\chi^2+116 M_W^4
M_\chi^4+\frac{1376}{3} M_\chi^6 M_W^2-1600 M_\chi^8\right)
\nonumber\\&
-2 E_e M_\chi^2\left(M_W^8 M_\chi^2-4 M_W^6 M_\chi^2+400 M_W^4 M_\chi^4+960 M_\chi^6
M_W^2-2^{10} M_\chi^8\right)
+2^8 M_\chi^3 \left(M_W^2+2 M_\chi^2\right)\bigg)
\nonumber\\&+2^8 \left(2 M_\chi^2+M_W^2-\frac{4 E_e^2 (M_\chi-E_e)^2}{M_W^2}\right)
\ln\left[\frac{M_W^2 M_\chi}{(M_\chi-E_e) \left(M_W^2+4 E_e
M_\chi\right)}\right]\Bigg\}+\Order(v^2,M_\eta^{-2},m_\ell^2).\label{dsigmadEl}
\end{align}
\end{widetext}

\section{Approximate Spectrum for Boosted {\boldmath $W$} Decay Products}
\label{app:approxWdk}

If any possible polarization of the produced $W$ is neglected, then a simple calculation results for the 
spectra of the finals state particles from $W$ decay.  The lab frame spectra of the decay product 
(of type or ``flavor'' $F$) depends on a one-dimensional convolution of the isotropic spectrum 
in the $W$ rest frame (RF energy $E'$), $\frac{dN_F}{dE'_F,}$,
with the $W$ spectrum in the lab frame, $\frac{dN}{dE_W}$.
We now develop this convolution.

Given the energy distribution $dN_W/d\gamma$ of produced $W$'s (with $\gamma=E_W/M_W$),
and the energy distribution $dN_F/dE'_F$ of decay particle $F$ in the $W$~rest frame,
normalized to the multiplicity of $F$ per $W$ decay 
(i.e., there is a branching ratio $W\rarr F$ multiplier implicit in $dN_F/dE'_F$)
and assumed to be isotropic,\footnote{
If the $W$ polarization is not neglected, then the $W$ decay amplitude includes Wigner functions
$d^1_{\mu_i \mu_f}(\theta)$, which introduce a linear $\cos\theta$ or $\sin\theta$ term into Eq.~\rf{Ezero}.
}
one gets the spectrum $dN_F/dE_F$ of particle $F$ in the lab via:
\bea{Ezero}
\frac{dN_F (E)}{dE}&=& \int_{-1}^1 \frac{d\cos\theta'}{2} \int d\gamma \frac{dN_W}{d\gamma} \\
     && \times \int dE'\,\frac{dN_F}{dE'}\ \delta(E-[\gamma E'+\beta\gamma p'\cos\theta'])\nonumber\,,
\eea
with $p'=\sqrt{E'^2-m^2_F}$, $\beta\gamma=\sqrt{\gamma^2-1}$.
The $\cos\theta'$ integral is easily done, and one gets 
\beq{E1}
\frac{dN_F (E)}{dE}=\half\int_1^\infty \frac{d\gamma}{\sqrt{\gamma^2-1}}\frac{dN_W}{d\gamma}
     \int_{E'_-}^{E'_+} \frac{dE'}{p'}\,\frac{dN_F}{dE'}\,,
\eeq
with $E'_\pm = \gamma E\pm\beta\gamma p$. Equivalently, we get 
\beq{E2}
\frac{dN_F (E)}{dE}=\half\int_{m_F}^\infty \frac{dE'}{p'}\,\frac{dN_F}{dE'}\,,\int_{\gamma_-}^{\gamma_
+} 
     \frac{d\gamma}{\sqrt{\gamma^2-1}}\frac{dN_W}{d\gamma}\,,
\eeq
with $\gamma_\pm=(EE'\pm pp')/m^2_F$ and $p=\sqrt{E^2-m^2_F}$.
This formulation neglects interferences between identical particles
produced in both the primary and secondary channels, if any.

As given, Eq.~\rf{E2} applies to any particle type in the $W$'s final state.
For example, it could be used to calculate the antiproton or antineutron spectrum from $W$ 
production and decay,
if the fragmentation functions for $W\rarr \bar{p} {\rm \ or\ }\bar{n}$, 
i.e. $f(x_{\bar{B}}\equiv 2E_{\bar{B}}/M_W)$ were input.

Here we perform a the convolution for the especially simple case of $W$ decay to
two massless particles,
say $\nu_\e$ and $e$.
For massless leptons, we have
\beq{massless2body}
\frac{dN_\nu}{dE'}=\frac{dN_\e}{dE'}=BR(W\rarr\nu\e)\,\delta(E'-\half M_W)\,,
\eeq
with
$\gamma_+=(E_W/M_W)_{max}=(s+M_W^2)/2\sqrt{s}M_W\approx (4M_\chi^2+M_W^2)/4M_\chi
M_W$,
and $\gamma_-=(4E^2+M_W^2)/4E M_W$.
The spectrum in the lab is given by Eq.~\rf{E2} becomes just
\beq{Emassless2body}
\frac{dN_\nu}{dE}=\frac{dN_e}{dE}= \frac{(BR)}{M_W}\, 
   \int_{\gamma_-}^{\gamma_+}
\frac{d\gamma}{\sqrt{\gamma^2-1}}\frac{dN_W}{d\gamma}\,.
\eeq
The $W$-spectrum shown in Fig.~\rf{dNdEW} is approximately half of an ellipse,
suggesting the fit
\beq{ellipse1}
\left(\frac{\ln\left(\frac{dN}{dx_W}\right)-\ln 0.07}{\ln 2.0 -\ln0.07}\right)^2 +
\left(\frac{x_W -0.65}{0.50\,(1.01-0.29)}\right)^2=1\,,
\eeq
valid for $0.29\alt x_W\alt 1.01$.
Solving for $dN/dx_W$ then gives
\beq{ellipse2}
\frac{dN}{dx_W}
= 0.07\,\left(\frac{2.0}{0.07}\right)^{\sqrt{ 1-7.7\,(x_W-0.65)^2}}\,.
\eeq
Substituting into Eq.~\rf{Emassless2body}
$\gamma=\frac{M_\chi}{M_W}\,x_W$, $\frac{dN}{d\gamma}=\frac{M_W}{M_\chi}\,\frac{dN}{dx_W}$,
and $\frac{dN}{dx_W}$ given in Eq.~\rf{ellipse2},
we obtain the desired one-dimensional integral for the secondary lepton spectrum, per
$W$:
\bea{ellipse3}
\frac{dN_\nu (x_\ell)}{dx_\ell}&=& 0.07\,(BR)\,
   \int_{x_-}^{x_+} \frac{dx_W}{\sqrt{x_W^2
-\left(\frac{M_W}{M_\chi}\right)^2}}\nonumber\\
&&\nonumber\\
&\ \times& \left(\frac{2.0}{0.07}\right)^{\sqrt{1-7.7\,(x_W-0.65)^2}}\,.
\eea
The integration limits are
$x_+=1+\frac{M_W^2}{4M_\chi^2}$,
and $x_-=\left(x_\ell+\frac{M_W^2}{4x_\ell\,M_\chi^2}\right)$.
The range of $x_\ell$ for the leptons from $W$~decay is
$[ \frac{M_W^2}{4M_\chi^2},\ 1 ]$. 
The relevant branching ratios~\cite{PDG} are
$BR(W\rarr \nu\e)=11\%$,
$BR(Z\rarr \nu\nubar)=6.7\%$,
and $BR(Z\rarr\ell^+\ell^-)=3.4\%$,
each per single flavor mode, $e,\ \mu,$ or $\tau$.
We show the resulting lepton spectrum, without the BR factor,
in Fig.~\rf{fig:SecLepSpectrum}.

\newpage



\begin{thebibliography}{99}




\bibitem{Kamionkowski_review}
  G.~Jungman, M.~Kamionkowski and K.~Griest,
  Phys.\ Rept.\  {\bf 267}, 195 (1996).

\bibitem{Bertone_review}
  G.~Bertone, D.~Hooper and J.~Silk,
  Phys.\ Rept.\  {\bf 405}, 279 (2005).

\bibitem{Bergstrom_review}
  L.~Bergstrom,
  Rept.\ Prog.\ Phys.\  {\bf 63}, 793 (2000)


  

\bibitem{hardgamma}  
  L.~Bergstrom,
  Phys.\ Lett.\  B {\bf 225}, 372 (1989);

  R.~Flores, K.~A.~Olive and S.~Rudaz,
  Phys.\ Lett.\  B {\bf 232}, 377 (1989);

E.~A.~Baltz and L.~Bergstrom,
  Phys.\ Rev.\  D {\bf 67}, 043516 (2003);
  [arXiv:hep-ph/0211325].

  L.~Bergstrom, T.~Bringmann and J.~Edsjo,
  Phys.\ Rev.\  D {\bf 78}, 103520 (2008)
  [arXiv:0808.3725 [astro-ph]];

 V.~Barger, Y.~Gao, W.~Y.~Keung and D.~Marfatia,
  Phys.\ Rev.\  D {\bf 80}, 063537 (2009)
  [arXiv:0906.3009 [hep-ph]].
  
\bibitem{Bringmann:2007nk}
  T.~Bringmann, L.~Bergstrom and J.~Edsjo,
  JHEP {\bf 0801}, 049 (2008).
  [arXiv:0710.3169 [hep-ph]].
  

\bibitem{Berezinsky:2002hq}
  V.~Berezinsky, M.~Kachelriess and S.~Ostapchenko,
  Phys.\ Rev.\ Lett.\  {\bf 89}, 171802 (2002) [arXiv:hep-ph/0205218].

\bibitem{Kachelriess}
  M.~Kachelriess and P.~D.~Serpico,
  Phys.\ Rev.\  D {\bf 76}, 063516 (2007) [arXiv:0707.0209 [hep-ph]].

\bibitem{BDJW}
  N.~F.~Bell, J.~B.~Dent, T.~D.~Jacques and T.~J.~Weiler,
  Phys.\ Rev.\  D {\bf 78}, 083540 (2008)
  [arXiv:0805.3423].
It has been pointed out in~\cite{KSS09} that the bremsstrahlung rate presented 
in this paper is too low by a factor of $\quarter$.  
This is so --- Eq.~(8) and the subsequent rate equations should be multiplied by~4.

\bibitem{Dent:2008qy}
  J.~B.~Dent, R.~J.~Scherrer and T.~J.~Weiler,
  Phys.\ Rev.\  D {\bf 78}, 063509 (2008)
  [arXiv:0806.0370 [astro-ph]].

\bibitem{Ciafaloni:2010qr}
  P.~Ciafaloni, A.~Urbano,
  Phys.\ Rev.\  {\bf D82}, 043512 (2010).
  [arXiv:1001.3950 [hep-ph]].


\bibitem{KSS09}
  M.~Kachelriess, P.~D.~Serpico and M.~A.~Solberg,
  arXiv:0911.0001 [hep-ph].

\bibitem{Chen:1998}
Massive three body final states were considered in X. l. Chen and M. Kamionkowski, JHEP {\bf 9807}, 001 (1998)


\bibitem{Pamela_positrons}
  O.~Adriani {\it et al.}  [PAMELA Collaboration],
  Nature {\bf 458}, 607 (2009)
  [arXiv:0810.4995 [astro-ph]].

\bibitem{Pamela_antiprotons}
  O.~Adriani {\it et al.},
  Phys.\ Rev.\ Lett.\  {\bf 102}, 051101 (2009)
  [arXiv:0810.4994 [astro-ph]].

\bibitem{:2010rc}
  O.~Adriani {\it et al.}  [PAMELA Collaboration],
  arXiv:1007.0821 [astro-ph.HE].

\bibitem{Fermi1}
  A.~A.~Abdo {\it et al.}  [The Fermi LAT Collaboration],
  Phys.\ Rev.\ Lett.\  {\bf 102}, 181101 (2009)
  [arXiv:0905.0025 [astro-ph.HE]].




\bibitem{pulsars}
  D.~Hooper, P.~Blasi and P.~D.~Serpico,
  JCAP {\bf 0901}, 025 (2009)
  [arXiv:0810.1527 [astro-ph]];

  H.~Yuksel, M.~D.~Kistler and T.~Stanev,
 Phys.\ Rev.\ Lett.\  {\bf 103}, 051101 (2009)
  [arXiv:0810.2784 [astro-ph]];

  S.~Profumo,
  arXiv:0812.4457 [astro-ph];

  D.~Malyshev, I.~Cholis and J.~Gelfand,
Phys.\ Rev.\  D {\bf 80}, 063005 (2009)
  [arXiv:0903.1310 [astro-ph.HE]];

  V.~Barger, Y.~Gao, W.~Y.~Keung, D.~Marfatia and G.~Shaughnessy,
Phys.\ Lett.\  B {\bf 678}, 283 (2009)
  [arXiv:0904.2001 [hep-ph]];

  D.~Grasso {\it et al.}  [FERMI-LAT Collaboration],
Astropart.\ Phys.\  {\bf 32}, 140 (2009)
  [arXiv:0905.0636 [astro-ph.HE]];

  P.~Mertsch and S.~Sarkar,
 Phys.\ Rev.\ Lett.\  {\bf 103}, 081104 (2009)
  [arXiv:0905.3152 [astro-ph.HE]];

  D.~Malyshev,
 JCAP {\bf 0907}, 038 (2009)
  [arXiv:0905.2611 [astro-ph.HE]];

  K.~Kashiyama, K.~Ioka and N.~Kawanaka,
  arXiv:1009.1141 [astro-ph.HE].


\bibitem{supernova}
  N.~J.~Shaviv, E.~Nakar and T.~Piran,
Phys.\ Rev.\ Lett.\  {\bf 103}, 111302 (2009)
  [arXiv:0902.0376 [astro-ph.HE]];

  Y.~Fujita, K.~Kohri, R.~Yamazaki and K.~Ioka,
 Phys.\ Rev.\  D {\bf 80}, 063003 (2009)
  [arXiv:0903.5298 [astro-ph.HE]].


\bibitem{accel}
  P.~Blasi,
Phys.\ Rev.\ Lett.\  {\bf 103}, 051104 (2009)
  [arXiv:0903.2794 [astro-ph.HE]];

  H.~B.~Hu, Q.~Yuan, B.~Wang, C.~Fan, J.~L.~Zhang and X.~J.~Bi,
  arXiv:0901.1520 [astro-ph];

  S.~Dado and A.~Dar,
  arXiv:0903.0165 [astro-ph.HE].


\bibitem{prop}
  L.~Stawarz, V.~Petrosian and R.~D.~Blandford,
 Astrophys.\ J.\  {\bf 710}, 236 (2010)
  [arXiv:0908.1094 [astro-ph.GA]];
R.~Cowsik and B.~Burch,
  arXiv:0905.2136 [astro-ph.CO];
  B.~Katz, K.~Blum and E.~Waxman,
  arXiv:0907.1686 [astro-ph.HE].

\bibitem{Fan:2010yq}
  Y.~Z.~Fan, B.~Zhang and J.~Chang,
  arXiv:1008.4646 [astro-ph.HE].



\bibitem{Randall}
  Y.~Cui, J.~D.~Mason and L.~Randall,
  arXiv:1006.0983 [hep-ph],
 and references therein.
 
\bibitem{Lindner:2010rr}
  M.~Lindner, A.~Merle and V.~Niro,
  arXiv:1005.3116 [hep-ph].

\bibitem{Kolb}
  M.~Beltran, D.~Hooper, E.~W.~Kolb and Z.~C.~Krusberg,
  Phys.\ Rev.\  D {\bf 80}, 043509 (2009)
  [arXiv:0808.3384 [hep-ph]].



\bibitem{Nishi:2004st}
  C.~C.~Nishi,
  Am.\ J.\ Phys.\  {\bf 73}, 1160 (2005)
  [arXiv:hep-ph/0412245].

\bibitem{IZp161-2}
C. Itzykson and J.B. Zuber, Quantum Field Theory, pages 161-2,
Dover Pr., 1980.

\bibitem{Taka1986}
  Y. Takahashi, {\sl ``The Fierz Identities''}, in {\sl Progress in Quantum Field Theory}, 
  ed. H. Ezawa and S. Kamefuchi (North Holland, Amsterdam, 1986), p. 121.


\bibitem{Peskin}
M. E. Peskin and D. V. Schroeder, An Introduction To Quantum Field Theory,
Westview Pr., 1995.


\bibitem{Weinberg}
  S.~Weinberg,
{\it  Cambridge, UK: Univ. Pr. (1995) 609 p}


\bibitem{Srednicki}
M. Srednicki, Quantum Field Theory,
Cambridge U. Press, 2007.


\bibitem{Haim1983}
H.~Goldberg,
  Phys.\ Rev.\ Lett.\  {\bf 50}, 1419 (1983),
making use of some earlier Fierzing by 
%
P.~Fayet,
 Phys.\ Lett.\  B {\bf 86}, 272 (1979).
%

A detailed calculation of the related amplitude $e^+ e^- \rarr
\photino\photino$ involving two identical Majorana particles is
available in App.\ E (as well as a lucid and complete presentation of
Feynman rules for Majorana fermions in App.\ D) of H.~E.~Haber and
G.~L.~Kane,
  Phys.\ Rept.\  {\bf 117}, 75 (1985).

Another lucid listing of Feynman rules for Majorana fermions is
available in Chapter 49 of ``Quantum Field Theory'', by M.~Srednicki,
Cambridge University Press.
  
See L.M.Krauss, Nucl.\ Phys.\ B {\bf 227} 556, (1983) for cosmological implications

\bibitem{Cao:2009yy}
  Q.~H.~Cao, E.~Ma and G.~Shaughnessy,
  Phys.\ Lett.\  B {\bf 673}, 152 (2009)
  [arXiv:0901.1334 [hep-ph]].


\bibitem{Ma:2000cc}
  E.~Ma,
  Phys.\ Rev.\ Lett.\  {\bf 86}, 2502 (2001)
  [arXiv:hep-ph/0011121].

\bibitem{Gelmini}
  G.~B.~Gelmini, E.~Osoba and S.~Palomares-Ruiz,
  Phys.\ Rev.\  D {\bf 81}, 063529 (2010)
  [arXiv:0912.2478 [hep-ph]].


\bibitem{Berezinsky}
P. ~Ciafaloni and D. ~Comelli, 
Phys.\ Lett.\ B {\bf{446}}, 278-284 (1999) [arXiv:hep-ph/9809321];

M. ~Beccaria, P.~Ciafaloni, D.~Comelli, F.M.~Renard, C.~Verzegnassi,
Phys.\ Rev.\ D {\bf{61}} 073005 (2000) [arXiv:hep-ph/9906319];

V.S.~Fadin, ~L.N. ~Lipatov, Alan D. ~Martin, M. ~Melles, 
Phys.\ Rev.\ D {\bf{61}} 094002 (2000) [arXiv:hep-ph/9910338];

W. ~Beenakker, A. ~Werthenbach,
Phys.\ Lett.\ B {\bf{489}} 148-156 (2000) [arXiv:hep-ph/0005316];

Michael ~Melles, 
Phys.\ Rept. {\bf{375}} 219-326 (2003) [arXiv:hep-ph/0104232];

  G.~Bell, J.~H.~Kuhn and J.~Rittinger,
  arXiv:1004.4117 [hep-ph];

  P.~Ciafaloni, D.~Comelli, A.~Riotto, F.~Sala, A.~Strumia and A.~Urbano,
  arXiv:1009.0224 [hep-ph];

and ref.~\cite{Berezinsky:2002hq}.


\bibitem{LL}
L.D. Landau and E.M. Lifschitz, ``The Classical Theory of Fields'',
Pergamon Press, $4^{th}$ revised Enlish edition, pages 32-34.

\bibitem{GelmGondo}
P. Gondolo and G. Gelmini, Nucl. Phys. B360, 145 (1991).


\bibitem{rapidity}
Some may recognize the logarithmic factor in Eq.~\rf{dsigdEW} as twice the 
rapidity of the produced $W$, i.e. $y_W=\half\ln(\frac{E_W+p_W}{E_W-p_W})\,.$


\bibitem{Crocker:2010gy}
  R.~M.~Crocker, N.~F.~Bell, C.~Balazs and D.~I.~Jones,
  Phys.\ Rev.\  D {\bf 81}, 063516 (2010)
  [arXiv:1002.0229 [hep-ph]].



\bibitem{anti-deuteron}
  S.~Schael {\it et al.}  [ALEPH Collaboration],
  Phys.\ Lett.\  B {\bf 639}, 192 (2006)
  [arXiv:hep-ex/0604023].

\bibitem{BDJWnext}
  N.~F.~Bell, J.~B.~Dent, T.~D.~Jacques and T.~J.~Weiler, 
in preparation.


\bibitem{PDG}
Summary Tables from the Particle Data Group,
http://pdg.lbl.gov/



\bibitem{Okun:1982ap}
  L.~B.~Okun,
  {\it Leptons And Quarks},
  Amsterdam, Netherlands, North-Holland (1982) 361p, section 29.3.5.
  
\bibitem{revisited}
  N.~F.~Bell, J.~B.~Dent, A.~J.~Galea, T.~D.~Jacques, L.~M.~Krauss, T.~J.~Weiler,
  [arXiv:1104.3823 [hep-ph]].


\end{thebibliography}
\end{document}